\documentclass[lettersize,journal]{IEEEtran}
\usepackage{amsmath,amsfonts}
\usepackage{array}
\usepackage[caption=false,font=normalsize,labelfont=sf,textfont=sf]{subfig}
\usepackage{textcomp}
\usepackage{stfloats}
\usepackage{url}
\usepackage{verbatim}
\usepackage{graphicx}
\usepackage{cite}
\usepackage{amssymb}
\usepackage{mathtools}
\usepackage{booktabs}
\usepackage{multirow}
\usepackage{tabularx}
\usepackage{hyperref}
\usepackage{color}
\usepackage{soul}
\usepackage[normalem]{ulem}
\usepackage[utf8]{inputenc}
\usepackage{amsmath}
\usepackage[ruled,vlined,linesnumbered]{algorithm2e}
\usepackage{setspace} 
\usepackage[table]{xcolor}

\newcommand{\orcid}[1]{\href{https://orcid.org/#1}{\includegraphics[width=8pt]{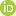}}}
\begin{document}

\title{EmoSphere++: Emotion-Controllable Zero-Shot Text-to-Speech via Emotion-Adaptive Spherical Vector}

\author{Deok-Hyeon~Cho\orcid{0009-0002-4673-9882},
        Hyung-Seok~Oh\orcid{0000-0001-7229-8123},
        Seung-Bin~Kim\orcid{0000-0002-2287-9111},
        and~Seong-Whan~Lee\orcid{0000-0002-6249-4996},~\IEEEmembership{Fellow,~IEEE}
\thanks{This work was partly supported by Institute of Information \& Communications Technology Planning \& Evaluation (IITP) grant funded by the Korea government (MSIT) (No. RS-2019-II190079, Artificial Intelligence Graduate School Program (Korea University), No. RS-2021-II-212068, Artificial Intelligence Innovation Hub, and No. RS-2024-00336673, AI Technology for Interactive Communication of Language Impaired Individuals). \textit{(Corresponding author: Seong-Whan Lee.)}}
\thanks{ D.-H. Cho, H.-S. Oh, S.-B. Kim and S.-W. Lee are with the Department of Artificial
Intelligence, Korea University, 145, Anam-ro, Seongbuk-gu, Seoul 02841, Republic of Korea.\protect
E-mail: (dh\_cho@korea.ac.kr\; hs\_oh@korea.ac.kr\; sb-kim@korea.ac.kr\; sw.lee@korea.ac.kr).}
}

\markboth{}%
{Shell \MakeLowercase{\textit{et al.}}: A Sample Article Using IEEEtran.cls for IEEE Journals}

\IEEEpubid{}
\IEEEpubidadjcol

\maketitle

\begin{abstract}
Emotional text-to-speech (TTS) has advanced significantly, but challenges persist due to the complexity of emotions and limitations in emotional speech datasets and models. A key issue with previous studies is the reliance on limited emotional speech datasets or extensive manual annotations, which restrict generalization across different speakers and emotional styles. To address this, we propose EmoSphere++, an emotion-controllable zero-shot TTS model capable of generating expressive speech with fine-grained control over emotional style and intensity—without requiring manual annotations. We introduce a novel emotion-adaptive spherical vector that effectively captures emotional style and intensity, along with a joint attribute style encoder that enhances generalization to both seen and unseen speakers. To further improve emotion transfer in zero-shot scenarios, we introduce an additional disentanglement method to enhance the style transfer performance for zero-shot scenarios. Through both objective and subjective evaluations, we demonstrate the benefits of the proposed model in emotion style and intensity modeling, as well as its effectiveness in enhancing emotional expressiveness across both seen and unseen speakers.
\end{abstract}


\section{Introduction}
\IEEEPARstart {E}{motions} are interrelated in a highly systematic fashion \cite{russell1980circumplex}. For example, the emotion of sadness can be expressed with derivative states of primary emotions such as feeling hurt or lonely, depending on the style and intensity. In speech synthesis, the ability to generate expressive and controllable emotional speech is essential for creating natural and effective human-computer interactions, as emotions are nuanced and can manifest in varying styles and intensities. Recently, emotional text-to-speech (TTS) technology has experienced rapid developments, increasing the interest in global interpretable emotion control \cite{wu2024laugh, cho24_interspeech, qi2024towards, zheng2024controllable}. Controllable emotional TTS represents a breakthrough in reproducing human-like emotions in speech synthesis, thus enabling more emotionally intelligent interactions between humans and computers. Although researchers have made significant progress in controlling emotional intensity, the ability to precisely control emotional style remains a challenge.

Modeling diverse emotional styles and intensities is a major challenge in controllable emotional TTS. Unlike discrete emotion categories, emotional style and intensity are highly subjective and complex, making them difficult to accurately represent. Two general approaches for achieving controllable emotional TTS involve controlling conditioning features or manipulating internal emotion representations. That is, one approach uses conditioning features of emotion intensity, such as relative ranking matrices \cite{zhu2019controlling, zhou2022emotion, lei2021fine, lei2022msemotts, zhou2023speech, inoue2024hierarchical}, distance-based quantization \cite{im2022emoq}, or voiced, unvoiced, and silence (VUS) states \cite{matsumoto2020controlling}. The alternative approach involves the manipulation of internal emotion representations through the application of scaling factors \cite{li22h_interspeech, li2022cross} or interpolation of the embedding space \cite{um2020emotional}. However, despite these methods, the explicit control of emotion style and intensity remains a largely unexplored topic in emotional speech synthesis.

\begin{figure}[t] 
    \centering
    \includegraphics[width=1.0\linewidth]{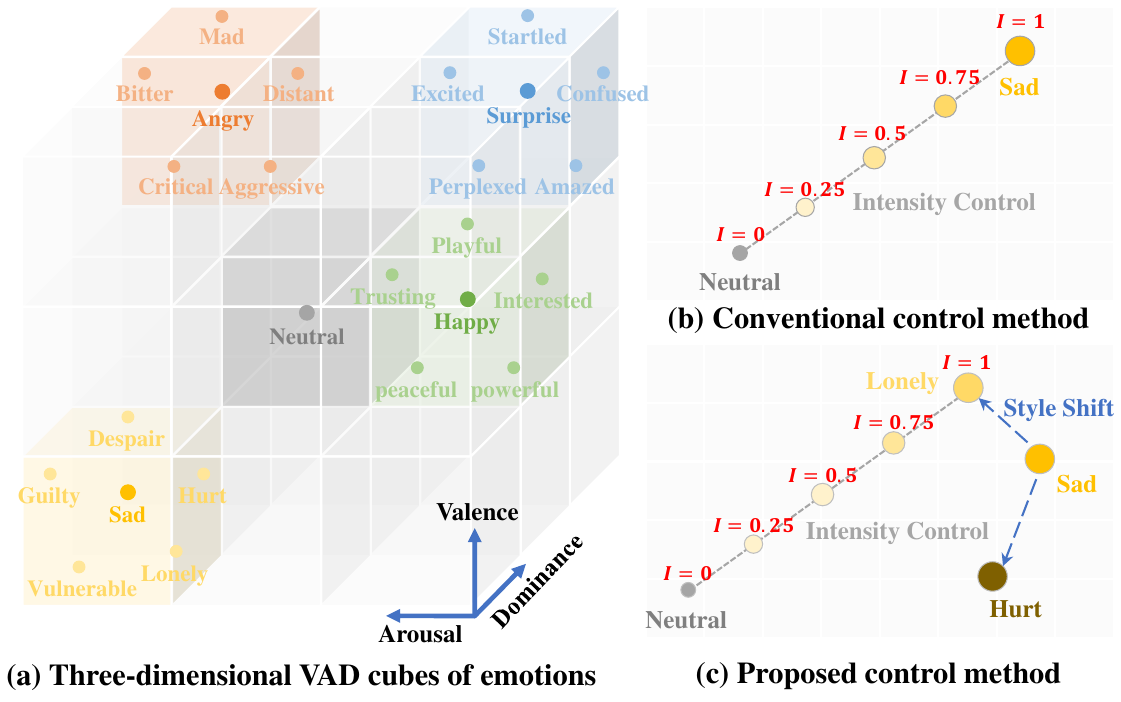}\vspace{-0.2cm}
    \caption{(a) Three-dimensional valence-arousal-dominance (VAD) cubes of emotions, where all emotional styles occur as derivative states of primary emotions. Emotional intensity control method is used for (b) conventional models and (c) the proposed model with consideration for emotional style.
} \vspace{0cm}
    \label{Emotion_control}
\end{figure}

Another approach to controlling emotional expression involves utilizing emotional dimensions. Compared to the discrete emotion approach, the dimensional approach, such as Russell's circumplex model, provides a more precise method for capturing the nuances between different emotional states \cite{russell1980circumplex}. Recently, studies on TTS systems have attempted to control emotional attributes through the emotion dimension \cite{habib2019semi, sivaprasad21_interspeech}. In one of these studies, a prosody control block is extended by incorporating the continuous space of arousal and valence to allow interpretable emotional prosody control \cite{sivaprasad21_interspeech}. Another study proposes an expressive TTS model with a semi-supervised latent variable to control emotions in six discrete emotional states of arousal-valence combinations \cite{habib2019semi}. However, this setup requires labor-intensive annotations, which are more expensive to obtain than categorical labels and more susceptible to annotator bias. The emotional dimension model also exhibits limitations when explicitly controlling emotion style and intensity.
To address these challenges, EmoSphere-TTS \cite{cho24_interspeech} models derivative emotions through emotional attribute prediction and discrete emotion labels, enabling explicit control over emotion style and intensity. However, limitations persist due to reliance on predefined emotion and speaker labels.

Most emotional TTS systems utilize Sequence-to-sequence (Seq2Seq) models, which not only predict the duration of speech automatically but also learn feature mapping and alignment simultaneously \cite{robinson2019sequence, kim2020emotional}. The attention mechanism in these models allows them to focus on the emotionally emphasized parts of an utterance \cite{yang22t_interspeech}. However, Seq2Seq models face the typical challenges of auto-regressive models, such as long-term dependence and repetition problems. Furthermore, most emotional speech syntheses adopt fine-tuning to control emotion intensity on a single speaker dataset; however, some of these methods exhibit noticeably degraded speech quality \cite{zhou2022emotion, zhou2023speech}. 
Researchers have explored acoustic models and additional discriminators for emotion transfer to enhance the capture of expressiveness when synthesizing acoustic features \cite{oh2024durflex, oh2024diffprosody, cho24_interspeech, zhou2024emotional}. 
However, existing methods primarily focus on emotion transfer using discrete emotion labels, which overlook the complexity of emotions conveyed in human speech \cite{li2022cross, pan2021cross}. Furthermore, these approaches often rely on additional discriminators to enhance expressiveness, adding complexity to the model while struggling to fully capture emotional nuance. To address these challenges, a TTS system capable of generalizing across zero-shot emotion transfer scenarios is needed, ensuring more accurate emotion synthesis without relying on predefined labels \cite{mehta2024matcha, kim2024p, le2024voicebox, eskimez2024e2, wu2024laugh}.

As previously discussed, existing study and prior work \cite{cho24_interspeech} face the following challenges: (1) Defining and modeling emotional style and intensity as derivatives of primary emotions, while accounting for characteristics such as the distribution of emotion categories; (2) Effectively integrating global and fine-grained emotion representations to enhance emotional expressiveness, while ensuring robust generalization across unseen speakers and emotions; (3) Designing a TTS system capable of achieving high generalization and expressive capability in zero-shot style transfer scenarios, without relying on additional modules; and (4) Evaluating synthesized emotional speech beyond global emotion assessment, to include subjective measurement of detailed emotion styles. To build upon these discussions, we were inspired by psychological studies \cite{plutchik2013theories, reisenzein1994pleasure} that have explored frameworks and methods for measuring complex emotions that arise from more primary emotional states. Additionally, with the increasing demand for personalized speech generation, we aimed to address the challenges in TTS models by focusing on achieving high generalization capability and producing high-quality speech. In this article, we present the following key contributions:
\begin{itemize}
    \item We introduce an emotion-adaptive coordinate transformation that models the emotion-adaptive spherical vector (EASV), enabling more interpretable and controllable synthesis of emotion style and intensity.
    \item We introduce a joint attribute style encoder along with an additional disentanglement module, enabling the model to perform emotion transfer even in zero-shot scenarios where reference speakers and emotions are not explicitly labeled.
    \item We propose a novel objective evaluation method, spherical vector angle similarity (SVAS), to evaluate overall emotion accuracy while also capturing subtle variations in speech emotion styles with greater precision.
    \item We carefully designed objective and subjective evaluations to demonstrate the effectiveness and contributions of the proposed model from multiple perspectives.
\end{itemize}

Audio samples and source code are available at \url{https://github.com/Choddeok/EmoSpherepp}.

\section{Background and Related Work}
\subsection{Characterization of Emotions}
The processes of defining and expressing emotions have garnered significant interest in psychology \cite{whissell1989dictionary, ekman1992argument, russell1980circumplex, schroder2006expressing}. Emotion theorists divide emotion theory into discrete \cite{ekman1986new, plutchik2013theories} and dimensional models \cite{russell1980circumplex, russell1977evidence}. Discrete models represent emotions as distinct, separate categories, while dimensional models provide a continuous and fine-grained description, capturing the complexity and variability of emotional experiences.

Emotion labels correspond very closely to the categories we use in our daily lives. Paul Ekman \cite{ekman1986new} derived six primary emotions: happiness, anger, disgust, sadness, anxiety, and surprise based on universally recognized facial expressions. However, this approach overlooks the nuanced variations of emotions. For instance, Plutchik's emotion wheel \cite{plutchik2013theories} proposes eight primary emotions and suggests that all other emotions arise as derivative states of the primary emotions. By adjusting the intensity of primary emotions on the wheel, a broader spectrum of emotional experiences can be represented. Although people can explicitly express emotions, modeling the relationships between discrete emotional states presents a challenge.

Researchers have introduced dimensional models to computationally interpret the relationships between emotional states. 
Dimensional models represent emotions along three continuous dimensions \cite{russell1980circumplex, russell1977evidence}: valence represents the positivity or negativity of an emotion, arousal indicates the intensity of the emotion provoked by a stimulus, and dominance denotes the level of control exerted by the stimulus. 
Russell's circumplex model \cite{russell1980circumplex} suggests a two-dimensional circular space that spans the independent and bipolar dimensions of arousal and valence. Building on this, researchers have attempted to extend the model to the third dimension of dominance to denote the location of emotion within this space \cite{russell1977evidence}. In the valence-arousal space, intensity is often equated with arousal; moreover, Reisenzein demonstrated that using the angle and length of the vector in polar coordinates is the only possible option for interpreting the relationships between emotions \cite{reisenzein1994pleasure, jenke2018cognitive}. Recently, the speech processing domain has seen various studies on emotion recognition that leverage the emotional dimension \cite{yang2023disentangled, yang2023cluster, xiao2019mes, shepstone2016audio}. 

Despite these efforts in psychology, the current literature on speech synthesis is still insufficient in effectively modeling and controlling the subtle variations of emotions. Inspired by several psychological theories \cite{plutchik2013theories, reisenzein1994pleasure}, we hypothesize that the dimensional model allows speech synthesis models to express derived emotions of primary emotions, as shown in Fig. \ref{Emotion_control} (a). This approach enables researchers to generate, control, and manipulate a wide range of emotions more easily in real-life applications.

\begin{figure*}[t] 
    \centering
    \includegraphics[width=1\linewidth]{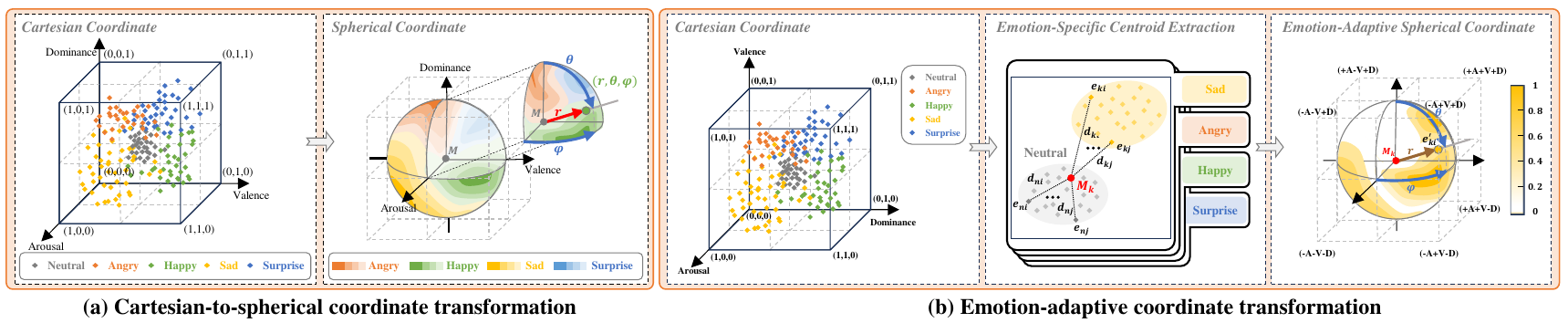}\vspace{-0.2cm}
    \caption{Illustration of coordinate transformations: (a) Cartesian-to-spherical coordinate transformation \cite{cho24_interspeech} and (b) Emotion-adaptive coordinate transformation. In (a), the center $M$ represents the central coordinates of the neutral state, whereas (b) introduces $M_{k}$, which serves as the representative center that reflects the distributions of both neutral and target emotions.}
    \label{Emotion-Adaptive_CT}
    \vspace{0cm}
\end{figure*}

\subsection{Controllable Emotional Speech Synthesis}
Recently, speech synthesis models have exhibited significant developments \cite{oh2024diffprosody, 10381805, 8341805}; therefore, controllable emotional speech synthesis research is being aggressively pursued \cite{qi2024towards, cho24_interspeech, zheng2024controllable}. Researchers typically use the following controllable emotional speech synthesis methods to utilize general emotional datasets: 1) emotion label- and 2) reference-based approaches.

The emotion label-based approach aims to properly model conditioning input to reflect the complex nature of emotions. Researchers typically model emotion intensity using a learned ranking function \cite{parikh2011relative}, as employed in \cite{zhu2019controlling, zhou2022emotion, zhou2023speech, lei2021fine, lei2022msemotts, inoue2024hierarchical}. The ranking function \cite{parikh2011relative} seeks a ranking matrix based on the relationships between the dimensional-driven and different global emotional expressions using support vector machines. The model receives the emotional intensity of emotional samples as a conditioning input for training. However, this method tends to rely on emotion labels and introduces bias into training through separate stages. Most models that use ranking functions adopt fine-tuning to control emotion intensity on a single-speaker dataset; however, some of these methods have noticeably degraded speech quality \cite{zhou2022emotion, zhou2023speech}. Moreover, certain research studies have utilized conditioning input such as distance-based quantization \cite{im2022emoq} and VUS states \cite{matsumoto2020controlling} to model emotional intensity. However, these methods are still limited to several predefined emotion labels and lack differentiation among samples within the same emotion label.

As emotional speech synthesis often lacks multiple emotional style labels, reference-based approaches are famous for using reference audio to transfer emotional styles. Several studies have controlled emotion intensity through operations on representative emotion embedding. The scaling factors approach \cite{li22h_interspeech, li2022cross} reflects fine-grained emotion representation through multiplication. In addition to the scaling approach, the interpolation approach proposed by \cite{um2020emotional} controls emotion intensity through an inter-to-intra emotional distance ratio algorithm. Despite these techniques, the structure of the embedding space influences model performance and complicates the process of finding optimal parameters for scaling or interpolation.

However, these methods cannot be tuned explicitly like label-based approaches, nor do they capture the fine-grained emotion representations achievable by reference-based methods. 
EmoSphere-TTS \cite{cho24_interspeech} solves this by proposing a spherical emotion vector to control the emotional style and intensity of the synthetic speech. However, the lack of consideration for emotion category distribution in emotion style and intensity modeling can lead to unnatural variations in certain styles and intensities. This study addresses the lack of emotion category distribution-based modeling by explicitly modeling emotion style and intensity variations based on EmoSphere-TTS \cite{cho24_interspeech}, thereby bridging this research gap.

\subsection{Style and Emotion Transfer in Text-to-Speech}
The TTS research community has long explored methods for style and emotion transfer. A significant advancement came with the introduction of global style tokens (GSTs) \cite{wang2018style}, which provided a framework for capturing and transferring speaking styles. One such derivative is Mellotron \cite{valle2020mellotron}, an autoregressive multi-speaker TTS model based on Tacotron that utilizes GSTs. Li et al. \cite{li2022cross} proposed a module for disentangling and transferring emotions across different speakers to achieve a desired emotional tone while maintaining the identity of the speaker. However, these methods exhibit limitations in capturing different styles and zero-shot scenarios owing to the restricted speaker lookup table and the insufficient performance of the disentangling modules.

In response, iEmoTTS \cite{zhang2023iemotts} and YourTTS \cite{casanova2022yourtts} used pre-trained speaker embedding for robust zero-shot performance. Additionally, GenerSpeech \cite{huang2022generspeech} proposed a multi-level style adapter to obtain different styles, including a global latent representation with speaker and emotion features. However, previous research has lacked methods to effectively process style and disentangle speech factors. Our work handles well-formed style processing through joint attribute style encoders and incorporation of additional loss to exhibit strong generalization performance for zero-shot scenarios.

\subsection{Speaker and Emotion Feature Disentanglement Methods}
Some prosodic features are inherently associated with the speaker's identity, making complete disentanglement challenging. Therefore, an effective disentangling method is critical to adequately separate speaker identity from emotion-related prosodic features, ensuring clear and accurate emotional transfer without compromising the target speaker's timbre. Researchers typically implement the disentanglement module via explicit labels, such as the gradient reversal layer (GRL) \cite{ganin2016domain}, as employed in \cite{zhang2023iemotts, oh2024durflex}. However, using explicit labels for disentanglement introduces a trade-off between preserving emotional information and separating speaker identity, making hyperparameter optimization challenging and leading to suboptimal performance and synthesis quality \cite{li2022cross}. Vector quantization (VQ) \cite{van2017neural} offers an alternative approach to separating information without relying on explicit labels. Although VQ is effective in separating information without explicit labels \cite{zhang2023iemotts, huang2022generspeech}, it often results in unintended information loss and requires complex optimization to balance compression with reconstruction quality. To address these issues, \cite{li2022cross} proposed an orthogonal loss to compensate for the embedding of emotion for the loss of emotional information caused by the disentanglement of speaker information \cite{ranasinghe2021orthogonal}. Building on these advancements, we aim to minimize information loss during disentanglement using an orthogonal loss-based approach while enhancing the expressiveness of synthesized speech.

\section{Proposed Method}
This paper introduces EmoSphere++, an emotion-controllable zero-shot TTS model that can control emotional style and intensity to resemble natural human speech. Our work is grounded in the emotion wheel theory \cite{plutchik2013theories}, which suggests that all other emotions are derived from the primary emotions, and the circumplex model \cite{russell1980circumplex, jenke2018cognitive}, representing emotions through coordinate transformations based on intensity. We propose that derivative emotions can be characterized by the coordinate transformation in the valence-arousal-dominance (VAD) dimension, reflecting the style and intensity of the primary emotions, as shown in Fig. \ref{Emotion_control} (c). 

Building on this, EmoSphere-TTS \cite{cho24_interspeech} was previously introduced to model emotion style and intensity using a spherical emotion vector. However, the lack of consideration for emotion category distribution can lead to unnatural variations in certain styles and intensities. To address this, we propose an emotion-adaptive coordinate transformation that better models diverse emotional styles and intensities. Additionally, we introduce a joint attribute style encoder to enable emotion-controllable zero-shot TTS across a broader range of emotions, overcoming the limitations of relying on predefined emotion and speaker labels, which restrict flexibility in emotional expression. Furthermore, we achieve competitive performance solely through a conditional flow matching (CFM)-based decoder, eliminating the need for an additional discriminator module while enhancing emotional expressiveness and speech quality. 
The details of our approach are defined in the following subsections.

\begin{figure}[t] 
    \centering
    \includegraphics[width=1\linewidth]{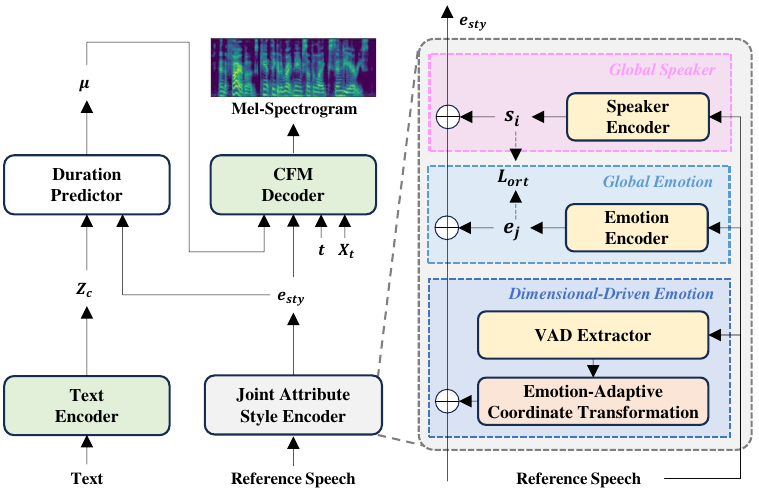}\vspace{-0.2cm}
    \caption{
    Training diagram of the EmoSphere++ framework. 
    The framework consists of three main modules: the text encoder, the joint attribute style encoder, and the conditional flow matching (CFM) decoder. The right section illustrates the detailed structure of the joint attribute style encoder, which extracts global speaker, global emotion, and dimensional-driven emotion to form a joint attribute style embedding for emotional speech synthesis.
} \vspace{0cm}
    \label{Model_overall}
\end{figure}

\subsection{Emotion-Adaptive Coordinate Transformation}
\label{Emotion-Adaptive Coordinate Transformation}
Several studies \cite{im2022emoq, zhou2022emotion, zhou2023speech, cho24_interspeech} assume that emotional intensity decreases when approaching a neutral state and use this as the basis for modeling emotion intensity. As shown in Fig. \ref{Emotion-Adaptive_CT} (a), previous studies \cite{cho24_interspeech} defined the center coordinates based on this assumption, where the intensity of emotion decreases as it approaches the neutral emotion center $M$, formulated as follows:
\begin{equation}
M = \frac{1}{N_{n}} \sum_{i=1}^{N_{n}} e_{i}^{n},
\end{equation}
where $N_{n}$ is the total number of neutral coordinates $e_{i}^{n}$. However, since the mean-based method does not account for the distribution of other emotions, such as variance, it fails to fully capture the relationship between the neutral emotion and the target emotion.
Specifically, we define other emotions as all emotions except the neutral emotion and the target emotion as a specific emotion selected from these other emotions.  
To address this, our approach simultaneously considers the distributions of the neutral emotion and the corresponding target emotion, extracting an adaptive spherical vector for each target emotion.  

Our method models a spherical coordinate system for each target emotion by considering the distribution of the neutral emotion and its corresponding target emotion, as shown in Fig. \ref{Emotion-Adaptive_CT} (b). 
Our approach is based on two fundamental assumptions: (1) the emotional intensity increases as it moves farther from the center of the emotion-adaptive spherical coordinate system and (2) the angle from the center of the emotion-adaptive spherical coordinate determines the emotional style. Initially, we adopted a specific emotional attribute prediction model \cite{wagner2023dawn} $\psi$ to predict the VAD value $e_{i}^{k}$ in emotion class $k$:
\begin{equation}
\label{eq2}
e_{i}^{k}=\psi(x_i),
\end{equation}
where $x_i$ denote the $i$-th referece speech in speech dataset $\mathrm{X}$ and $e_{i}^{k}$ consists of values for $(d_{v},d_{a},d_{d})$, where $d_{v}$, $d_{a}$, and $d_{d}$ represent valence, arousal, and dominance, respectively. Each component is expressed in Cartesian coordinates, with values ranging from 0 to 1. To model the spherical coordinate system for each emotion, we obtained the shifted Cartesian coordinates $\widehat{e_{i}^{k}}=(\widehat{d}_{v}, \widehat{d}_{a}, \widehat{d}_{d})$ by shifting through the representative central coordinates $M_{k}$ from different target emotion coordinate set $E_{k}$. The coordinates $M_{k}$ are extracted using emotion-specific centroid extraction, which maximizes the ratio of the distance between specific target emotions to the distance from the neutral coordinates as follows:

\begin{algorithm}[!t]
\caption{Emotion-Adaptive Coordinate Transformation}\label{alg}
\setstretch{1.2}
    \KwIn{Emotional speech dataset $\mathrm{X}$, SER model $\psi$}
    
    \KwOut {Emotion-adaptive spherical vector set $\mathbb{S}$}
    
    \textbf{for} (reference speech $x_i$, emotion class $k$) in $\mathrm{X}$ \textbf{do}:
    
        \quad Get VAD value $e_{i}^{k} \in E_k$ using SER model $\psi$ by Eq. (\ref{eq2})

    \textbf{end for}

    \textbf{for} each emotion class $k$ of $E_k$ \textbf{do}:

        \quad \textbf{if} $E_k$ is the neutral class set $E_n$ \textbf{then}:

        \quad\quad \textbf{for} $e_{i}^{n}$ in $E_n$ \textbf{do}:

        \quad\quad\quad Append $s_{i}^{n} = (0, 0, 0)$ to $\mathbb{S}$

        \quad\quad \textbf{end for}

        \quad \textbf{else}:

        \quad\quad Compute centroid coordinate as $M_k$ by Eq. (\ref{eq3})

        \quad\quad \textbf{for} $e_{i}^{k}$ in $E_k$ \textbf{do}:
    
            \quad\quad\quad Compute shifted VAD as $\widehat{e_{i}^{k}}$ by Eq. (\ref{eq4})
            
            \quad\quad\quad Spherical transformation as $s_{i}^{k} \in S_{k}$ by Eq. (\ref{eq5})
        
        \quad\quad \textbf{end for}

        \quad\quad Calculate ($r_{min}$, $r_{max}$) the interquartile range of $r$

        \quad\quad \textbf{for} $s_{i}^{k}$ in $S_{k}$ \textbf{do}:
              
            \quad\quad\quad Compute $r_{IQR}$ by Eq. (\ref{eq6})

            \quad\quad\quad Append $s_{i}^{k} = (r_{IQR}, \vartheta, \varphi)$ to $\mathbb{S}$

        \quad\quad \textbf{end for}
    
    \textbf{end for}

\end{algorithm}

\begin{equation}
\label{eq3}
M_{k}=\arg\max_{M}\frac{\mathbb{E}_{e_{i}^{k} \in E_{k}}[ \| M - e_{i}^{k} \|_2 ]}{\mathbb{E}_{e_{i}^{n} \in E_{n}}[\|M-e_{i}^{n}\|_2]},
\end{equation}
\begin{equation}
\label{eq4}
\widehat{e_{i}^{k}} = e_{i}^{k} - M_{k}.
\end{equation}

Here $e_{i}^{k}$ and $e_{i}^{n}$ denote the $i$-th coordinate of the $k$-th target emotion coordinate set $E_{k}$ and the neutral coordinate set $E_{n}$, respectively. Consequently, the centroid coordinates are the ones that maximize the distances from the target emotion category while minimizing the distance within the neutral emotion category. Then, transformation via the representative central coordinates $M_{k}$ to spherical coordinates $s_{i}^{k}=(r, \vartheta, \varphi)$ can be formulated as follows:

\begin{equation*}
r = \sqrt{{\widehat{d_v}}^{2} + {\widehat{d_a}}^{2} + {\widehat{d_d}}^{2}},
\end{equation*}
\begin{equation}
\label{eq5}
\vartheta = \arccos\left(\frac{\widehat{d}_{d}}{r}\right),
\varphi = \arctan\left(\frac{\widehat{d}_{v}}{\widehat{d}_{a}}\right).
\end{equation}

After the emotion-adaptive coordinate transformation, we applied the interquartile range (IQR) technique \cite{walfish2006review} to adjust the data based on the median, thereby reducing the influence of outliers as follows:

\begin{equation*}
r_{clamp} = \min(\max(r, r_{min}), r_{max}),
\end{equation*}
\begin{equation}
\label{eq6}
r_{IQR} = \frac{r_{clamp} - r_{min}}{r_{max} - r_{min}}.
\end{equation}

Here $r_{min}$ and $r_{max}$ are bounds derived based on the IQR technique with $r_{min}$ set as the first quartile minus 1.5 times the IQR and $r_{max}$ as the third quartile plus 1.5 times the IQR. The detailed procedure for obtaining the emotion-adaptive spherical vector set $\mathbb{S}$ is outlined in Algorithm \ref{alg}.

\subsection{Joint Attribute Style Encoder}
The voice typically contains highly dynamic style attributes (e.g. speaker identities, prosody and emotions), making the TTS model difficult to model and transfer in a zero-shot scenario. As shown in Fig. \ref{Model_overall}, we propose a joint attribute style encoder for both broad and fine-grained stylization.

The emotion encoder includes a fine-tuned categorical emotion recognition model for global emotion features and an additional EASV extractor for dimensional-driven emotion features. 
The global emotion encoder extracts the fixed-size hidden embedding from the categorical emotion recognition\footnote{\url{https://github.com/ddlBoJack/emotion2vec}} \cite{ma2023emotion2vec}, which utilizes a multilayer transformer to capture comprehensive emotional representations. 
Meanwhile, the EASV extractor, described in Section \ref{Emotion-Adaptive Coordinate Transformation}, generates dimensional-driven emotion features, which explicitly encode fine-grained variations in emotional expressions (e.g., intensity and style). Finally, a fully connected layer processes global and dimensional-driven emotion features, combining them into a fixed-size hidden embedding.

The speaker encoder provides speaker-related information to the TTS model. A pretrained speech encoder extracts a speaker embedding \cite{jia2018transfer} from the reference speech for zero-shot emotion transfer. We used the WavLM Base model \cite{chen2022wavlm} as the speaker verification model\footnote{\url{https://huggingface.co/microsoft/wavlm-base-sv}}, building the model on the HuBERT \cite{hsu2021hubert} framework to focus on modeling speech content and preserving speaker identity. WavLM-based approaches can capture and represent speaker-specific information and emotional nuances in speech \cite{chakhtouna2023statistical, yang2024single}. Similar to emotion embeddings, fixed-size hidden embeddings are processed in the fully connected layer and combined to generate joint attribute style embedding $e_{sty}$.

\subsection{Preliminary on Conditional Flow Matching-Based Model}
\label{Preliminary on Conditional Flow Matching-Based Model}
EmoSphere++ adopts a CFM-based decoder that generates flow through the ordinary differential equation. Building on the success of flow matching in the speech synthesis task \cite{mehta2024matcha, le2024voicebox, kim2024p}, we utilized a CFM-based decoder that is designed to model a conditional vector field $\mathbf{u}_t$. Following \cite{lipmanflow}, we define the flow $\phi$ as the mapping between two density functions:

\begin{equation}
\tfrac{d}{dt}\phi_t(x) = \boldsymbol{v}_t(\phi_t(x)) \text{;}
\quad\quad
\phi_0(x) = x \text{.}
\end{equation}

Here, $\boldsymbol{v}_t$ represents a time-dependent vector field that defines the path of the probability flow over time $t \in \left [0, 1 \right]$. Specifically, it describes a conditional flow process in which the conditional flow $\phi_{t,x1}$ represents simple linear trajectories between the data point $x_{1}$ drawn from the target distribution $q(x)$ and prior distribution $x_{0} \sim N(0, I)$:

\begin{equation}
\phi_{t,x1}(x_{0}) = (1 - (1 - \sigma_{min})t)x_{0} + tx_{1}, 
\end{equation}
where $\sigma_{min}$ is the hyper-parameter for small amounts of white noise. The vector field in the decoder is trained using the following objectives:

{\scriptsize 
\begin{equation}
\mathit{L}_{cfm} = \mathbb{E}_{t,q(x_{1}),p(x_{0})}\left\| u_{t}(\phi_{t,x_{1}}(x_{0}))-\widetilde{\upsilon}_{\theta}(\phi_{t,x_{1}}(x_{0}),\mu,e_{sty},t) \right\|^{2},
\tag*{\normalsize (8)}
\end{equation}}
where $\mu$ represents the predicted average acoustic features (e.g., Mel-spectrogram) given the text and the chosen durations, using a text encoder and duration predictor. $e_{sty}$ denotes the joint attribute style embedding.

\subsection{Training Objective}
Alongside traditional losses for training TTS systems, we introduce a disentanglement method to help in modeling our system to control and transfer emotion style and intensity.

First, text encoder and duration predictor architectures were implemented following \cite{mehta2024matcha}. Duration-model training uses monotonic alignment search to compute the duration loss $L_{dur}$ and the prior loss $L_{enc}$, as described in \cite{popov2021grad}. The decoder follows the CFM loss $L_{cfm}$ described in Section \ref{Preliminary on Conditional Flow Matching-Based Model}.

Inspired by \cite{li2022cross}, we introduced an additional orthogonality loss of the disentanglement method.
In the style encoder, emotion and speaker embeddings contain overlapping information regarding each other. The speaker and emotion embeddings should be 1) discriminative in distinguishing identities and 2) independent of each other, ensuring effective generalization performance for both seen and unseen speakers. To mitigate the impact of speaker and emotion embedding leakage on model performance, we propose a normalized orthogonality loss $L_{ort}$ to enhance the decoupling capability of the model.
Unlike the existing loss \cite{li2022cross} applied only to embeddings from the same audio, this method normalizes all sample pairs to enhance generalization in zero-shot scenarios. In this case, $L_{ort}$ can be expressed as:

\begin{equation}
L_{ort}= \sum_{j=1}^{n}\sum_{i=1}^{n}\frac{\left\|s_{i}^{T}e_{j} \right\|^{2}}{\left\|s_{i}\right\|^{2}\left\|e_{j}\right\|^{2}},
\end{equation}
where $n$ is the batch size, and the Frobenius norm $\left\| \cdot \right\|$ is used to calculate the interaction between the emotion embedding $e_{i}$ and speaker embedding $s_{j}$ for all pairs of samples $(i,j)$.

Consequently, the final objective function is defined as:
\begin{multline}
    L_{total} = \lambda_{enc}L_{enc} + \lambda_{cfm}L_{cfm} + \lambda_{dur}L_{dur} + \lambda_{ort}\mathcal{L}_{ort},
\end{multline}
where $\lambda_{enc}$, $\lambda_{cfm}$, $\lambda_{dur}$, and $\lambda_{ort}$ are the loss weights, which we set to 1.0, 1.0, 1.0, and 0.02, respectively. 

\begin{figure}[t] 
    \centering
    \includegraphics[width=1\linewidth]{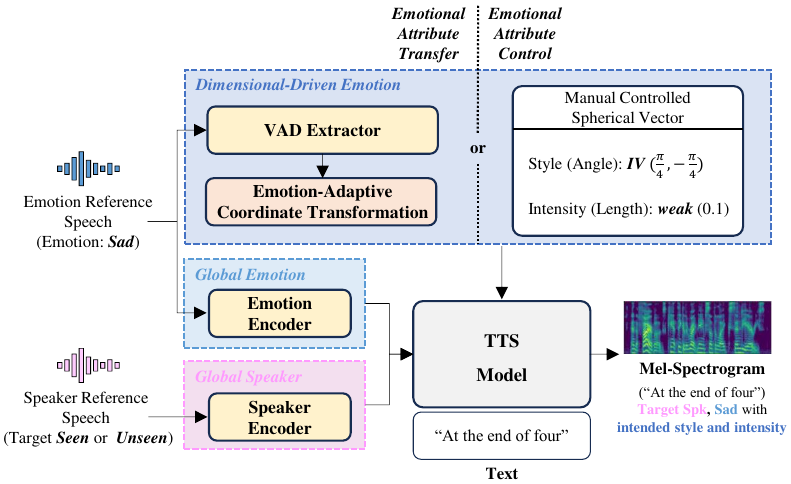}\vspace{-0.2cm}
    \caption{Run-time diagram of the proposed EmoSphere++ framework. We can manually control the emotion style and intensity via the dimensional-driven emotion of emotion style and intensity. We produce an emotional state as the derivative of primary emotions by assigning the appropriate angle and length to the spherical vector.
} \vspace{0cm}
    \label{run-time}
\end{figure}

\subsection{Control of Emotional Style and Intensity}
Fig. \ref{run-time} illustrates the proposed emotion-controllable zero-shot TTS framework, which synthesizes emotional speech for both seen and unseen speakers based on reference speech and EASV. The framework comprises three main modules: the text encoder, joint attribute style encoder, and CFM decoder.

The joint attribute style encoder captures the emotion and speaker information in an embedding from the reference speech. By varying the length and angle in the EASV, we can manipulate the levels of style and intensity at runtime and efficiently synthesize the desired emotional effects. In a zero-shot scenario, the framework can synthesize speech for unseen speakers by inputting the unseen target reference speech into the speaker encoder.

\section{Experiments}
\subsection{Experimental Setup}
We conducted experiments using the emotional speech dataset (ESD)\footnote{\url{https://github.com/HLTSingapore/Emotional-Speech-Data}}\cite{ZHOU20221}, which contains 350 parallel utterances spoken by ten English speakers in five emotional states (neutral, happy, angry, sad, and surprise). Following the prescribed data partitioning criteria, we extracted one sample for each emotion from every speaker, resulting in 17,500 samples. The validation set comprised 20 samples for each emotion per speaker, totaling 1,000 samples, whereas the test set comprised 30 samples for each emotion per speaker, totaling 1,500 samples. The zero-shot scenario used two unseen speakers, one English-speaking male (``0013'') and one English-speaking female (``0019''), by excluding them from the training process. 

Moreover, we utilized the MSP-Podcast corpus dataset \cite{lotfian2017building} and the Interactive Emotional Dyadic Motion Capture (IEMOCAP) dataset \cite{busso2008iemocap}, along with the ESD dataset, to analyze the prosodic variation of EASV and to verify whether the models can reflect the styles using predicted VAD values. The MSP-Podcast corpus dataset comprises approximately 237 hours of speech data annotated with both categorical emotion labels and dimensional VAD values. The training set includes eight categorical emotion classes (happiness, sadness, fear, surprise, contempt, disgust, and neutral) collected from 454 speakers. For dimensional emotion labels, raters evaluated VAD using a seven-point Likert scale. The IEMOCAP contains ten speakers with nine emotions (angry, excited, fear, sad, surprised, frustrated, happy, disappointed, and neutral) and dimensional labels such as VAD. For analyzing the prosodic variation of EASV, we used the MSP-Podcast and IEMOCAP datasets while ensuring consistency with ESD by selecting only five categorical emotion labels that match those in the ESD dataset. Additionally, we used the IEMOCAP dataset to verify whether the models can reflect styles using predicted VAD values. The validation set included ten samples per speaker, while the test set comprised 15 samples per speaker.

\begin{table*}[!t]
\caption{
Prosodic variation analysis of emotion-adaptive spherical vector values 
across the ESD, IEMOCAP, and MSP-Podcast emotional speech datasets. 
R1, R2, and R3 represent intensity regions divided by 0.33 and 0.66 thresholds. Green and red colors indicate the highest and lowest prosodic values within each intensity level, respectively. 
The variation ranges Rc shows the difference between the highest and lowest prosodic values.
``-'' indicates 
data not applicable
, with neutral emotions fixed at intensity 0.
}\vspace{-0.5cm}
\label{Table1}
\begin{center}
\renewcommand{\arraystretch}{1.0}
\resizebox{1.0\textwidth}{!}
{
\begin{tabular}{l|c|ccc|c|ccc|c|c|ccc|c|c|ccc|c|c}
\toprule
    \multirow{2}{*}{\textbf{Emotion}} & \multirow{2}{*}{\textbf{Style}} & \multicolumn{4}{c|}{\textbf{Num.}} & \multicolumn{5}{c|}{\textbf{Pitch (mean)}} & \multicolumn{5}{c|}{\textbf{Energy (mean)}} & \multicolumn{5}{c}{\textbf{Duration (mean)}} \\
\cmidrule{3-21} 
    & & \textbf{R1} & \textbf{R2} & \textbf{R3} & \textbf{All} & \textbf{R1} & \textbf{R2} & \textbf{R3} & \textbf{Rc} & \textbf{AVG} & \textbf{R1} & \textbf{R2} & \textbf{R3} & \textbf{Rc} & \textbf{AVG} & \textbf{R1} & \textbf{R2} & \textbf{R3} & \textbf{Rc} & \textbf{AVG} \\
\midrule
    \multirow{1}{4em}{\centering\arraybackslash \textbf{Neutral}}   &  \cellcolor{gray!10}\textbf{All (Average)} & \cellcolor{gray!10} - & \cellcolor{gray!10} - & \cellcolor{gray!10} - & \cellcolor{gray!10} 41,845 & \cellcolor{gray!10} - & \cellcolor{gray!10} - & \cellcolor{gray!10} - & \cellcolor{gray!10} - & \cellcolor{gray!10} 48.7 & \cellcolor{gray!10} - & \cellcolor{gray!10} - & \cellcolor{gray!10} - & \cellcolor{gray!10} - & \cellcolor{gray!10} 3.2 & \cellcolor{gray!10} - & \cellcolor{gray!10} - & \cellcolor{gray!10} - & \cellcolor{gray!10} - & \cellcolor{gray!10} 3.7 \\
\midrule
    \multirow{8}{4em}{\centering\arraybackslash \textbf{Angry}}     &  I (+V +A +D) &  611 & 2,558 & 480 & 3,649 & \cellcolor{red!10} 66.5 &  72.9 & \cellcolor{green!10} 79.0 &  12.5 &  72.5 & \cellcolor{red!10} 6.0 &  7.7 & \cellcolor{green!10} 10.9 &  4.9 &  7.8 & \cellcolor{green!10} 4.1 &  4.0 & \cellcolor{red!10} 3.7 &  0.4 &  4.0 \\
\cmidrule{2-21}
                                                                    &  III (-V -A +D) &  529 &  2,675 &  1,123 &  4,327 & \cellcolor{green!10} 59.7 &  56.1 & \cellcolor{red!10} 41.7 &  18.0 &  54.4 & \cellcolor{green!10} 4.8 &  2.8 & \cellcolor{red!10} 0.8 &  4.0 &  2.7 & \cellcolor{green!10} 3.9 &  3.0 & \cellcolor{red!10} 1.8 &  2.1 &  2.8 \\
\cmidrule{2-21}
                                                                    &  IV (+V -A +D) &  169 &  144 &  6 &  319 & \cellcolor{green!10} 71.1 &  69.7 & \cellcolor{red!10} 64.9 &  6.2 &  70.4 & \cellcolor{green!10} 8.9 &  6.4 & \cellcolor{red!10} 5.3 &  3.6 &  5.9 & \cellcolor{green!10} 3.8 &  3.6 & \cellcolor{red!10} 2.6 &  1.2 &  3.7 \\
\cmidrule{2-21}
                                                                     & V (+V +A -D) &  729 &  3,702 &  923 &  5,354 & \cellcolor{red!10} 66.1 &  72.8 & \cellcolor{green!10} 82.6 &  16.5 &  73.5 & \cellcolor{red!10} 5.7 &  7.2 & \cellcolor{green!10} 15.1 &  9.4 &  8.3 & \cellcolor{green!10} 4.2 &  3.8 & \cellcolor{red!10} 3.8 &  0.4 &  3.8 \\
\cmidrule{2-21}
                                                                    &  VI (-V +A -D) &  126 &  107 &  7 &  240 & \cellcolor{green!10} 55.8 &  50.9 & \cellcolor{red!10} 38.0 &  17.8 &  53.4 & \cellcolor{green!10} 5.2 &  4.6 & \cellcolor{red!10} 3.4 &  1.8 &  4.9 & \cellcolor{green!10} 3.9 &  3.3 & \cellcolor{red!10} 2.4 &  1.5 &  3.6 \\
\cmidrule{2-21}
                                                                    & VII (-V -A -D) & 530 & 1,763 & 648 & 2,941 & \cellcolor{green!10} 58.1 & 54.0 & \cellcolor{red!10} 37.3 & 20.8 & 53.1 & \cellcolor{green!10} 4.8 & 2.9 & \cellcolor{red!10} 0.6 & 4.2 & 3.0 & \cellcolor{green!10} 3.9 & 2.5 & \cellcolor{red!10} 1.6 & 2.3 & 2.6 \\
\cmidrule{2-21}
                                                                    & VIII (+V -A -D) & 138 & 51 & - & 189 & \cellcolor{red!10} 69.9 & \cellcolor{green!10} 72.4 & - & 2.5 & 70.5 & \cellcolor{green!10} 5.5 & \cellcolor{red!10} 5.1 & - & 0.4 & 5.4 & \cellcolor{green!10} 3.9 & \cellcolor{red!10} 3.3 & - & 0.3 & 3.7 \\

\cmidrule{2-21}
                                                                    & \cellcolor{gray!10} \textbf{All (Average)} & \cellcolor{gray!10} 2,832 & \cellcolor{gray!10} 11,000 & \cellcolor{gray!10} 3,187 & \cellcolor{gray!10} 17,019 & \cellcolor{gray!10} 63.9 & \cellcolor{gray!10} 64.1 & \cellcolor{gray!10} 57.3 & \cellcolor{gray!10} - & \cellcolor{gray!10} 64.0 & \cellcolor{gray!10} 5.8 & \cellcolor{gray!10} 5.2 & \cellcolor{gray!10} 6.0 & \cellcolor{gray!10} - & \cellcolor{gray!10} 5.4 & \cellcolor{gray!10} 4.0 & \cellcolor{gray!10} 3.4 & \cellcolor{gray!10} 2.7 & \cellcolor{gray!10} - & \cellcolor{gray!10} 3.5 \\
\midrule
    \multirow{9}{4em}{\centering\arraybackslash \textbf{Happy}}     & I (+V +A +D) & 825 & 5,326 & 1,624 & 7,775 & \cellcolor{red!10} 58.9 & 64.1 & \cellcolor{green!10} 72.2 & 13.3 & 65.1 & \cellcolor{red!10} 5.2 & 6.2 & \cellcolor{green!10} 8.7 & 3.5 & 6.6 & \cellcolor{green!10} 4.4 & 4.1 & \cellcolor{red!10} 3.9 & 0.5 & 4.1 \\
\cmidrule{2-21}
                                                                    & II (-V +A +D) & 233 & 382 & 49 & 664 & \cellcolor{green!10} 54.0 & 49.7 & \cellcolor{red!10} 46.3 & 7.7 & 50.9 & \cellcolor{red!10} 3.5 & \cellcolor{green!10} 3.9 & 3.6 & 0.4 & 3.8 & 4.2 & \cellcolor{green!10} 4.3 & \cellcolor{red!10} 3.9 & 0.4 & 4.2 \\

\cmidrule{2-21}
                                                                    & III (-V -A +D) & 1,104 & 3,715 & 330 & 5,149 & \cellcolor{green!10} 52.5 & 46.7 & \cellcolor{red!10} 36.4 & 16.1 & 47.5 & \cellcolor{green!10} 4.0 & 3.1 & \cellcolor{red!10} 1.5 & 2.5 & 3.2 & \cellcolor{green!10} 4.6 & 4.3 & \cellcolor{red!10} 3.7 & 0.9 & 4.4 \\
\cmidrule{2-21}
                                                                    & IV (+V -A +D) & 328 & 360 & 6 & 694 & \cellcolor{green!10} 61.0 & 59.0 & \cellcolor{red!10} 43.7 & 17.3 & 59.9 & \cellcolor{green!10} 4.9 & 4.2 & \cellcolor{red!10} 2.8 & 2.1 & 4.5 & \cellcolor{green!10} 4.3 & 3.7 & \cellcolor{red!10} 4.0 & 0.3 & 4.0 \\

\cmidrule{2-21}
                                                                    & V (+V +A -D) & 627 & 2,757 & 1,568 & 4,952 & \cellcolor{red!10} 61.4 & 68.3 & \cellcolor{green!10} 74.4 & 13.0 & 68.7 & \cellcolor{red!10} 5.2 & 5.6 & \cellcolor{green!10} 6.6 & 1.4 & 5.8 & \cellcolor{green!10} 4.1 & 3.3 & \cellcolor{red!10} 2.4 & 1.7 & 3.2 \\
\cmidrule{2-21}
                                                                    & VI (-V +A -D) & 225 & 376 & 149 & 750 & \cellcolor{red!10} 54.0 & \cellcolor{green!10} 59.0 & 56.7 & 5.0 & 56.6 & \cellcolor{green!10} 3.4 & 2.5 & \cellcolor{red!10} 1.8 & 1.6 & 2.8 & \cellcolor{green!10} 3.5 & 2.2 & \cellcolor{red!10} 1.7 & 1.8 & 2.5 \\
\cmidrule{2-21}
                                                                    & VII (-V -A -D) & 866 & 4,306 & 1,527 & 6,699 & \cellcolor{green!10} 54.7 & 50.6 & \cellcolor{red!10} 42.4 & 12.3 & 49.9 & \cellcolor{green!10} 4.1 & 3.1 & \cellcolor{red!10} 1.7 & 2.4 & 3.0 & \cellcolor{green!10} 4.4 & 3.7 & \cellcolor{red!10} 2.9 & 1.5 & 3.6 \\
\cmidrule{2-21}
                                                                    & VIII (+V -A -D) & 231 & 168 & 18 & 417 & \cellcolor{red!10} 63.7 & 68.5 & \cellcolor{green!10} 69.5 & 5.8 & 65.6 & \cellcolor{green!10} 5.2 & 4.6 & \cellcolor{red!10} 1.9 & 3.3 & 4.9 & \cellcolor{green!10} 4.1 & 3.5 & \cellcolor{red!10} 1.7 & 2.4 & 3.7 \\
\cmidrule{2-21}
                                                                    & \cellcolor{gray!10} \textbf{All (Average)} & \cellcolor{gray!10} 4,439 & \cellcolor{gray!10} 17,390 & \cellcolor{gray!10} 5,271 & \cellcolor{gray!10} 27,100 & \cellcolor{gray!10} 57.5 & \cellcolor{gray!10} 58.2 & \cellcolor{gray!10} 55.2 & \cellcolor{gray!10} - & \cellcolor{gray!10} 58.0 & \cellcolor{gray!10} 4.4 & \cellcolor{gray!10} 4.2 & \cellcolor{gray!10} 3.6 & \cellcolor{gray!10} - & \cellcolor{gray!10} 4.3 & \cellcolor{gray!10} 4.2 & \cellcolor{gray!10} 3.6 & \cellcolor{gray!10} 3.0 & \cellcolor{gray!10} - & \cellcolor{gray!10} 3.7 \\
\midrule
\multirow{9}{4em}{\centering\arraybackslash \textbf{Sad}}           & I (+V +A +D) & 412 & 2,142 & 810 & 3,364 & \cellcolor{red!10} 47.7 & 52.7 & \cellcolor{green!10} 61.9 & 14.2 & 54.6 & \cellcolor{red!10} 2.0 & 2.8 & \cellcolor{green!10} 3.9 & 1.9 & 3.0 & \cellcolor{red!10} 3.5 & 3.9 & \cellcolor{green!10} 4.3 & 0.8 & 3.9 \\
\cmidrule{2-21}
                                                                    & II (-V +A +D) & 146 & 146 & 7 & 299 & 44.0 & \cellcolor{green!10} 44.4 & \cellcolor{red!10} 39.7 & 4.7 & 44.1 & \cellcolor{green!10} 1.3 & \cellcolor{green!10} 1.3 & \cellcolor{red!10} 1.1 & 0.2 & 1.3 & \cellcolor{green!10} 3.0 & 2.9 & \cellcolor{red!10} 2.1 & 0.9 & 2.9 \\
\cmidrule{2-21}
                                                                    & III (-V -A +D) & 464 & 1,882 & 285 & 2,631 & \cellcolor{green!10} 48.2 & 41.1 & \cellcolor{red!10} 30.8 & 17.4 & 41.3 & \cellcolor{green!10} 1.6 & 1.2 & \cellcolor{red!10} 1.0 & 0.6 & 1.2 & \cellcolor{red!10} 3.0 & 2.9 & \cellcolor{green!10} 3.1 & 0.1 & 2.9 \\
\cmidrule{2-21}
                                                                    & IV (+V -A +D) & 111 & 95 & 3 & 209 & 51.3 & \cellcolor{green!10} 51.4 & \cellcolor{red!10} 24.8 & 26.6 & 50.8 & 1.7 & \cellcolor{green!10} 2.1 & \cellcolor{red!10} 1.6 & 0.5 & 1.9 & \cellcolor{red!10} 3.0 & 3.5 & \cellcolor{green!10} 4.6 & 1.6 & 3.2 \\
\cmidrule{2-21}
                                                                    & V (+V +A -D) & 419 & 1,797 & 589 & 2,805 & \cellcolor{red!10} 50.9 & 56.1 & \cellcolor{green!10} 65.2 & 14.3 & 57.5 & \cellcolor{red!10} 2.4 & 3.0 & \cellcolor{green!10} 4.4 & 2.0 & 3.2 & \cellcolor{red!10} 3.1 & 3.3 & \cellcolor{green!10} 3.6 & 0.5 & 3.4 \\
\cmidrule{2-21}
                                                                    & VI (-V +A -D) & 120 & 132 & 7 & 259 & 42.6 & \cellcolor{green!10} 45.2 & \cellcolor{red!10} 36.0 & 9.2 & 43.8 & 1.3 & \cellcolor{green!10} 1.8 & \cellcolor{red!10} 0.3 & 1.5 & 1.5 & \cellcolor{green!10} 3.0 & 2.7 & \cellcolor{red!10} 1.8 & 1.2 & 2.8 \\
\cmidrule{2-21}
                                                                    & VII (-V -A -D) & 411 & 2,138 & 891 & 3,440 & \cellcolor{green!10} 48.8 & 43.5 & \cellcolor{red!10} 33.0 & 15.8 & 41.8 & \cellcolor{green!10} 1.6 & 1.1 & \cellcolor{red!10} 0.7 & 0.9 & 1.0 & \cellcolor{green!10} 2.8 & \cellcolor{red!10} 2.6 & 2.8 & 0.2 & 2.7 \\
\cmidrule{2-21}
                                                                    & VIII (+V -A -D) & 115 & 84 & 1 & 200 & \cellcolor{red!10} 53.2 & 63.9 & \cellcolor{green!10} 97.6 & 44.4 & 57.5 & \cellcolor{green!10} 1.3 & 1.5 & \cellcolor{red!10} 0.3 & 1.0 & 1.3 & \cellcolor{green!10} 2.7 & 2.4 & \cellcolor{red!10} 1.1 & 1.6 & 2.5 \\
\cmidrule{2-21}
                                                                    & \cellcolor{gray!10} \textbf{All (Average)} & \cellcolor{gray!10} 2,198 & \cellcolor{gray!10} 8,416 & \cellcolor{gray!10} 2,593 & \cellcolor{gray!10} 13,207 & \cellcolor{gray!10} 48.3 & \cellcolor{gray!10} 49.8 & \cellcolor{gray!10} 48.6 & \cellcolor{gray!10} - & \cellcolor{gray!10} 48.9 & \cellcolor{gray!10} 1.7 & \cellcolor{gray!10} 1.8 & \cellcolor{gray!10} 1.7 & \cellcolor{gray!10} - & \cellcolor{gray!10} 1.8 & \cellcolor{gray!10} 3.0 & \cellcolor{gray!10} 3.0 & \cellcolor{gray!10} 2.9 & \cellcolor{gray!10} - & \cellcolor{gray!10} 3.0 \\
\midrule 
\multirow{7}{4em}{\centering\arraybackslash \textbf{Surprise}}      & I (+V +A +D) & 258 & 946 & 423 & 1,627 & \cellcolor{red!10} 71.2 & \cellcolor{green!10} 72.3 & 71.3 & 1.1 & 71.9 & \cellcolor{red!10} 3.4 & 4.3 & \cellcolor{green!10} 6.6 & 3.2 & 4.8 & \cellcolor{red!10} 2.4 & 3.0 & \cellcolor{green!10} 3.4 & 1.0 & 3.0 \\
\cmidrule{2-21}
                                                                    & III (-V -A +D) & 233 & 1,231 & 283 & 1,747 & \cellcolor{green!10} 61.7 & 50.8 & \cellcolor{red!10} 41.9 & 19.8 & 50.5 & \cellcolor{green!10} 3.7 & 3.3 & \cellcolor{red!10} 2.7 & 1.0 & 3.2 & \cellcolor{red!10} 3.0 & 3.8 & \cellcolor{green!10} 4.1 & 1.1 & 3.7 \\
\cmidrule{2-21}
                                                                    & IV (+V -A +D) & 61 & 89 & 16 & 166 & 67.8 & \cellcolor{red!10} 64.4 & \cellcolor{green!10} 72.2 & 7.8 & 66.4 & 4.1 & \cellcolor{green!10} 5.0 & \cellcolor{red!10} 2.4 & 2.6 & 4.4 & \cellcolor{red!10} 3.0 & 3.5 & \cellcolor{green!10} 3.6 & 0.6 & 3.3 \\
\cmidrule{2-21}
                                                                    & V (+V +A -D) & 256 & 1,298 & 437 & 1,991 & \cellcolor{red!10} 71.4 & 79.9 & \cellcolor{green!10} 84.5 & 13.1 & 79.7 & \cellcolor{red!10} 3.0 & 3.6 & \cellcolor{green!10} 5.1 & 2.1 & 3.8 & \cellcolor{green!10} 2.4 & 2.3 & \cellcolor{red!10} 2.2 & 0.2 & 2.3 \\
\cmidrule{2-21}
                                                                    & VII (-V -A -D) & 252 & 1,052 & 229 & 1,533 & \cellcolor{green!10} 59.2 & 54.8 & \cellcolor{red!10} 45.1 & 14.1 & 54.4 & \cellcolor{green!10} 3.6 & 2.8 & \cellcolor{red!10} 1.6 & 2.0 & 2.7 & \cellcolor{green!10} 2.8 & 2.6 & \cellcolor{red!10} 2.4 & 0.4 & 2.6 \\
\cmidrule{2-21}
                                                                    & VIII (+V -A -D) & 63 & 71 & 5 & 139 & \cellcolor{red!10} 76.0 & \cellcolor{green!10} 89.9 & 84.0 & 13.9 & 82.7 & \cellcolor{green!10} 4.1 & \cellcolor{red!10} 2.3 & 3.6 & 1.8 & 3.2 & \cellcolor{green!10} 2.4 & 1.8 & \cellcolor{red!10} 1.6 & 0.8 & 2.1 \\
\cmidrule{2-21}
                                                                    & \cellcolor{gray!10} \textbf{All (Average)} & \cellcolor{gray!10} 1,123 & \cellcolor{gray!10} 4,687 & \cellcolor{gray!10} 1,393 & \cellcolor{gray!10} 7,203 & \cellcolor{gray!10} 67.9 & \cellcolor{gray!10} 68.7 & \cellcolor{gray!10} 66.5 & \cellcolor{gray!10} - & \cellcolor{gray!10} 67.6 & \cellcolor{gray!10} 3.7 & \cellcolor{gray!10} 3.6 & \cellcolor{gray!10} 3.7 & \cellcolor{gray!10} - & \cellcolor{gray!10} 3.7 & \cellcolor{gray!10} 2.7 & \cellcolor{gray!10} 2.8 & \cellcolor{gray!10} 2.9 & \cellcolor{gray!10} - & \cellcolor{gray!10} 2.8 \\
\bottomrule
\end{tabular}
}\vspace{+0cm}
\end{center}
\end{table*}

For the Mel-spectrogram, we transformed audio using the short-time Fourier transform with a hop size of 256, a window size of 1,024, an fast Fourier transform size of 1,024, and 80 Mel bins. We converted the text to phoneme using the grapheme-to-phoneme tool of the Festival Speech Synthesis System \cite{black2001festival} to serve as the input to the text encoder. We designed parallel and non-parallel test scenarios at runtime depending on whether the input text matches the reference speech.

\subsection{Implementation Details}
For the acoustic model, we followed the Matcha-TTS \cite{mehta2024matcha} configuration, utilizing the text encoder and duration predictor in the encoder, along with a 1D U-Net based CFM decoder. 
The CFM decoder consists of two downsampling blocks, followed by two midblock and two upsampling blocks, each containing a Transformer layer with a hidden dimensionality of 256, an attention module with dimensionality of 64, and ``snakebeta'' activations \cite{lee2023bigvgan}.
Following \cite{li2019neural}, the text encoder was modified by incorporating relative position representations and adding a residual connection to the encoder pre-net. Based on \cite{kim2020glow}, the duration predictor consists of two convolutional layers with rectified linear unit activation, followed by layer normalization, dropout, and a projection layer. Regarding the emotional attribute prediction\footnote{\href{https://huggingface.co/audeering/wav2vec2-large-robust-12-ft-emotion-msp-dim}{huggingface.co/wav2vec2-large-robust-12-ft-emotion-msp-dim}}, we adopt a system proposed in \cite{wagner2023dawn}, which predicts VAD using wav2vec 2.0 \cite{baevski2020wav2vec} and a linear predictor. In the joint attribute style encoder, the emotion and speaker global encoders utilize emotion2vec model  \footnote{\url{https://github.com/ddlBoJack/emotion2vec}} \cite{ma2023emotion2vec} and the WavLM-based speaker verification model\footnote{\url{https://huggingface.co/microsoft/wavlm-base-sv}}, respectively. During training, both global encoders are frozen and extract hidden embeddings, which are then passed through a two-layer fully connected network. We trained the generator using random segments of 32 frames from the Mel-spectrogram, with a batch size of 32 and a total of 11M training steps. The AdamW optimizer was used, with a $1\times 10^{-4}$ learning rate. In the inference stage, the guidance level $\gamma$ was set to 100. We trained the vocoder using the official BigVGAN\footnote{\url{https://github.com/NVIDIA/BigVGAN}}\cite{lee2023bigvgan} implementation, incorporating LibriTTS \cite{zen19_interspeech}, voice cloning toolkit\footnote{\url{https://datashare.ed.ac.uk/handle/10283/2651}}, and ESD datasets. All comparison models were trained using a single NVIDIA RTX A6000 GPU.

\subsection{Evaluation}
\subsubsection{Subjective Metrics}
We adopted two subjective metrics: 1) a mean opinion score (MOS) evaluation and 2) a preference test to evaluate emotion expressiveness. We conducted subjective evaluations using Amazon Mechanical Turk. All subjects were required to listen with headphones and replay each sample 2-3 times.

We conducted a MOS evaluation for naturalness (nMOS), speaker similarity (sMOS), and emotion similarity (eMOS) using a nine-point scale ranging from 1 to 5, with increments of 0.5 units. The results are presented with a confidence interval of 95\%. 
20 subjects evaluated the samples by assessing the full set of extracted samples, where two samples were randomly selected for each emotion and speaker combination from the entire test set of 100 samples (2 × 5 (\# of emotions) × 10 (8 seen speakers and 2 unseen speakers)), ensuring consistency across models.

We also conducted a preference test to evaluate the modeling of emotion expressiveness. To demonstrate the success of our modeling, we synthesized speech with three different levels of emotion intensity (weak, medium, and strong) from pairs of speech that share the same emotion and style. Evaluators were presented with two different sentences with the same emotion and style, each with varying intensities, and tasked with selecting the one exhibiting a stronger emotion. We uniformly referred to scores 0.1 as weak, 0.5 as medium, and 0.9 as strong. 
20 subjects evaluated the samples by assessing the full set of extracted paired samples, where two pairs (four samples) were randomly selected for each emotion and speaker combination from a total of 200 samples in the test set (2 × 2 (pairs) × 5 (\# of emotions) × 10 (8 seen speakers and 2 unseen speakers)), ensuring consistency across models.

\subsubsection{Objective Metrics}
To evaluate linguistic consistency, we calculated the word error rate (WER) using the Whisper large model \cite{radford2023robust} (WER\textsubscript{Whis}) or wav2vec model \cite{baevski2020wav2vec} (WER\textsubscript{w2v}). For WER\textsubscript{AVG}, we computed the average of the WER obtained from both the Whisper and wav2vec models. For the speaker similarity measurements, we calculated the speaker embedding cosine similarity via Resemblyzer\footnote{\url{https://github.com/resemble-ai/Resemblyzer}} (SECS\textsubscript{R}) or WavLM\footnote{\url{https://huggingface.co/microsoft/wavlm-base-sv}} (SECS\textsubscript{W}) between the target and converted speech. For SECS\textsubscript{AVG}, we computed the average of the speaker embedding cosine similarities obtained from both the Resemblyzer and WavLM models. For prosodic evaluation, we computed the root mean square error for both pitch error (RMSE$_{f_0}$) and periodicity error (RMSE$_{period}$), along with the F1 score of voiced/unvoiced classification (F1$_{v/uv}$). For emotionally expressive evaluation, we determined the emotion classification accuracy (ECA) using a prebuilt emotion classification model emotion2vec \cite{ma2023emotion2vec} and the emotion embedding cosine similarity (EECS) \cite{oh2024durflex} by computing the cosine similarity of the emotion2vec hidden emotion embedding between the synthesized audio and arbitrary reference audio with the target emotion. We used emotion2vec+ base\footnote{\url{https://github.com/ddlBoJack/emotion2vec}}, a pretrained model that supports nine classes, and only used the five sentiment classes in the ESD dataset for evaluation.
Moreover, we propose the spherical vector angle similarity (SVAS) to evaluate the emotion of synthesized speech. The SVAS is obtained by computing the cosine similarity of the angle of the emotion spherical vector between the synthesized audio and reference audio with the target emotion. We used an emotional attribute prediction model\footnote{\href{https://huggingface.co/audeering/wav2vec2-large-robust-12-ft-emotion-msp-dim}{huggingface.co/wav2vec2-large-robust-12-ft-emotion-msp-dim}} \cite{wagner2023dawn} to predict the VAD values and applied a Cartesian-to-spherical transformation through fixed neutral center coordinates to extract the angle of the emotion spherical vector. Unlike conventional metrics, SVAS not only captures global emotional information but also provides a fine-grained evaluation of subtle emotional variations, enabling a more comprehensive assessment of synthesized speech emotions.
All evaluations were conducted using the official test set of the ESD dataset\footnote{\url{https://github.com/HLTSingapore/Emotional-Speech-Data}}\cite{ZHOU20221} to ensure a standardized evaluation.

\subsection{Comparison Models}
\label{Comparison Models}
We compared the similarity and quality of samples generated using the proposed EmoSphere++ with those produced by other systems. We used the same vocoder and open code, and the comparative models are summarized as follows:
\begin{itemize}
\item \textbf{GT and BigVGAN \cite{lee2023bigvgan}}: The ground truth (GT) audio and waveforms are generated from the ground truth Mel-spectrogram using vocoder.
\item \textbf{Mellotron} \cite{valle2020mellotron}: This auto-regressive multi-speaker TTS model allows for direct style control by conditioning on rhythm and pitch utilizes conditioning GSTs.
\item \textbf{Mixedemotion} \cite{zhou2023speech}: A relative attribute ranking-based model that pre-computes intensity values for mixed emotion synthesis and allows manual control at run time.
\item \textbf{YourTTS} \cite{casanova2022yourtts}: A zero-shot multi-speaker TTS with the pre-trained speaker encoder. Unlike other comparison models, it does not use a pre-trained vocoder through end-to-end training.
\item \textbf{GenerSpeech} \cite{huang2022generspeech}: A high-fidelity zero-shot style transfer method for out-of-domain TTS. We used the emotion2vec model instead of the fine-tuned wav2vec 2.0 model to capture global style for a fair comparison.
\item \textbf{iEmoTTS} \cite{zhang2023iemotts}: A non-autoregressive TTS model for a cross-speaker emotion transfer system developed based on timbre-prosody disentanglement.
\item \textbf{EmoSphere++}: Our proposed CFM-based emotion-controllable zero-shot TTS with EASV.
\end{itemize}

\begin{table*}[!ht]
    \centering 
        \caption{Subjective and objective evaluation results for non-parallel emotion transfer on the seen dataset.\\ The nMOS, sMOS, and eMOS scores are presented with 95\% confidence intervals.}
    \label{Table2}\vspace{-0.2cm}
        \resizebox{1.0\textwidth}{!}{
    \begin{tabular}{l|ccc|ccc|ccc|ccc}
        \toprule
        \multirow{2}{*}{\textbf{Method}} & \multicolumn{3}{c|}{\textbf{Subjective Evaluation}} & \multicolumn{8}{c}{\textbf{Objective Evaluation}} \\ 
        \cmidrule{2-13} 
         & \textbf{nMOS} ($\uparrow$) & \textbf{sMOS} ($\uparrow$) & \textbf{eMOS} ($\uparrow$) & \textbf{WER\textsubscript{Whis}} ($\downarrow$) & \textbf{WER\textsubscript{w2v}} ($\downarrow$) & \textbf{WER\textsubscript{AVG}} ($\downarrow$) & \textbf{SECS\textsubscript{R}} ($\uparrow$) & \textbf{SECS\textsubscript{W}} ($\uparrow$) & \textbf{SECS\textsubscript{AVG}} ($\uparrow$) & \textbf{ECA} ($\uparrow$) & \textbf{SVAS} ($\uparrow$) & \textbf{EECS} ($\uparrow$) \\ 
        \midrule
            GT & 4.06$\pm$0.05 & 4.15$\pm$0.05 & 4.11$\pm$0.05 & 11.42 & 14.99 & 13.21 & 0.7358 & 0.9264 & 0.8311 & 95.53 & - & 0.9487 \\ 
            BigVGAN \cite{lee2023bigvgan} & 3.95$\pm$0.05 & 4.01$\pm$0.06 & 3.98$\pm$0.06 & 11.36 & 15.15 & 13.26 & 0.7271 & 0.9231 & 0.8251 & 94.25 & 0.9815 & 0.9389 \\ 
        \midrule
            Mellotron \cite{valle2020mellotron} & 3.42$\pm$0.07 & 3.95$\pm$0.07 & 3.75$\pm$0.07 & \textbf{14.49} & 18.90 & \textbf{16.70} & 0.6981 & 0.8939 & 0.7960 & 46.93 & 0.7838 & 0.5061 \\
            Mixedemotion \cite{zhou2023speech} & 3.36$\pm$0.07 & 3.89$\pm$0.07 & 3.69$\pm$0.07 & 20.24 & 26.48 & 23.36 & 0.6920 & 0.8764 & 0.7842 & 63.44 & 0.8669 & 0.6282 \\
        \midrule
            YourTTS \cite{casanova2022yourtts} & 3.52$\pm$0.06 & 3.94$\pm$0.06 & 3.75$\pm$0.07 & 28.71 & 32.79 & 30.75 & 0.7296 & 0.6751 & 0.6811 & 74.84 & 0.8657 & 0.7890 \\
            Generspeech \cite{huang2022generspeech} & 3.85$\pm$0.06 & 3.96$\pm$0.06 & \textbf{3.86$\pm$0.06} & 16.63 & 22.56 & 19.60 & 0.7074 & 0.8888 & 0.7981 & 82.54 & 0.8460 & 0.8366 \\
            iEmoTTS \cite{zhang2023iemotts} & 3.77$\pm$0.06 & 3.94$\pm$0.06 & 3.79$\pm$0.07 & 26.07 & 29.74 & 27.91 & 0.6153 & 0.8208 & 0.7181 & 77.60 & 0.8000 & 0.7474 \\
        \midrule
            EmoSphere ++ (Proposed) & \textbf{3.92$\pm$0.06} & \textbf{3.97$\pm$0.06} & \textbf{3.86$\pm$0.06} & 15.52 & \textbf{18.85} & 17.19 & \textbf{0.7314} & \textbf{0.9047} & \textbf{0.8181} & \textbf{93.53} & \textbf{0.8717} & \textbf{0.9270}\\ 
        \bottomrule
    \end{tabular}
      }\vspace{0cm}
\end{table*}

\begin{table*}[!ht]
    \centering 
        \caption{Subjective and objective evaluation results for non-parallel emotion transfer on the unseen dataset. \\ The nMOS, sMOS, and eMOS scores are presented with 95\% confidence intervals.}
    \label{Table3}\vspace{-0.2cm}
        \resizebox{1.0\textwidth}{!}{
    \begin{tabular}{l|ccc|ccc|ccc|ccc}
        \toprule
        \multirow{2}{*}{\textbf{Method}} & \multicolumn{3}{c|}{\textbf{Subjective Evaluation}} & \multicolumn{8}{c}{\textbf{Objective Evaluation}} \\ 
        \cmidrule{2-13} 
         & \textbf{nMOS} ($\uparrow$) & \textbf{sMOS} ($\uparrow$) & \textbf{eMOS} ($\uparrow$) & \textbf{WER\textsubscript{Whis}} ($\downarrow$) & \textbf{WER\textsubscript{w2v}} ($\downarrow$) & \textbf{WER\textsubscript{AVG}} ($\downarrow$) & \textbf{SECS\textsubscript{R}} ($\uparrow$) & \textbf{SECS\textsubscript{W}} ($\uparrow$) & \textbf{SECS\textsubscript{AVG}} ($\uparrow$) & \textbf{ECA} ($\uparrow$) & \textbf{SVAS} ($\uparrow$) & \textbf{EECS} ($\uparrow$) \\ 
        \midrule
            GT & 4.02$\pm$0.05 & 4.04$\pm$0.05 & 4.08$\pm$0.05 & 13.06 & 15.15 & 14.11 & 0.7725 & 0.9282 & 0.8504 & 100.00 & - & 0.9944 \\ 
            BigVGAN \cite{lee2023bigvgan} & 3.93$\pm$0.06 & 4.01$\pm$0.05 & 4.04$\pm$0.06 & 13.11 & 17.07 & 15.09 & 0.7603 & 0.9275 & 0.8439 & 99.32 & 0.9844 & 0.9903 \\ 
        \midrule
            YourTTS \cite{casanova2022yourtts} & 3.67$\pm$0.07 & 3.93$\pm$0.05 & 3.71$\pm$0.08 & 26.48 & 32.78 & 29.63 & \textbf{0.6870} & 0.8220 & 0.7545 & 61.46 & 0.8323 & 0.6509 \\
            Generspeech \cite{huang2022generspeech} & 3.90$\pm$0.06 & 3.96$\pm$0.06 & 3.83$\pm$0.07 & 15.42 & 23.46 & 19.44 & 0.6812 & 0.8191 & 0.7502 & 80.57 & 0.8131 & 0.7935 \\
            iEmoTTS \cite{zhang2023iemotts} & 3.83$\pm$0.06 & 3.93$\pm$0.05 & 3.82$\pm$0.08 & 26.74 & 32.70 & 29.72 & 0.5970 & 0.7480 & 0.6725 & 51.03 & 0.7615 & 0.5364 \\
        \midrule
            EmoSphere ++ (Proposed) & \textbf{3.91$\pm$0.06} & \textbf{3.99$\pm$0.05} & \textbf{3.85$\pm$0.07} & \textbf{14.41} & \textbf{18.43} & \textbf{16.42} & 0.6543 & \textbf{0.8640} & \textbf{0.7592} & \textbf{94.61} & \textbf{0.8458} & \textbf{0.9358} \\ 
        \bottomrule
    \end{tabular}
      }\vspace{0cm}
\end{table*}

To ensure a fair comparison between the proposed method and existing emotion intensity modeling approaches, we trained the models using the same dataset and configurations as Matcha-TTS \cite{mehta2024matcha}, including the encoder (i.e., text encoder and duration predictor) and the U-Net-based decoder. We replaced the speaker and emotion attribute control module for a contrastive study with two competing controllable emotion methods: scaling factor and relative attributes through comprehensive experiments.

\begin{itemize}
    \item \textbf{Matcha-TTS w/ Scaling Factor}: Here, the emotion embedding, extracted through the emotion disentangling module, is multiplied by a scaling factor to control the emotion strength \cite{li2022cross}. The system manages the speaker identity controller using speaker embeddings obtained from a speaker look-up table.
    \item \textbf{Matcha-TTS w/ Relative Attributes}: Here, the relative attributes vector is obtained from a ranking function to control the emotion strength \cite{zhou2022emotion}. We employed the same emotion encoder and speaker look-up table, except for fine-tuning the emotion encoder with a single-speaker emotion dataset.
\end{itemize}

In summary, we conducted emotion transfer experiments to compare EmoSphere++ with other systems, including Mellotron, Mixedemotion, YourTTS, GenerSpeech, and iEmoTTS. Moreover, we evaluated emotion intensity control by comparing it with other intensity control methods, such as Matcha-TTS w/ Scaling Factor and Matcha-TTS w/ Relative Attributes.

\section{Results}
\subsection{Analysis of Emotion-Adaptive Spherical Vectors for Emotional Style and Intensity}
\label{Analysis of Emotional Style and Intensity Modeling}
As previously mentioned, we characterized the derivative states of emotions through emotion style and intensity using the emotion-adaptive spherical vector (EASV). To evaluate the effectiveness of modeling based on emotion style and intensity, we conducted an analysis using the ESD dataset along with large-scale datasets, including MSP-Podcast corpus \cite{lotfian2017building} and IEMOCAP datasets \cite{busso2008iemocap}. In this study, we divided the emotion space into eight regions (``I''$\sim$``VIII'') based on the VAD axes and then shifted the style along spherical spaces, using only the five categorical emotion labels from the ESD dataset for consistency. As shown in Table \ref{Table1}, we illustrate the prosodic variation and distribution of large-scale emotional speech datasets based on emotion style and intensity modeling. The analysis details are described in the following subsections.

\subsubsection{Analysis of Emotion Dataset Distribution by Style and Intensity}
To demonstrate the diverse modeling of emotions, we analyzed the distribution of emotion styles and intensities using EASV modeling. We observe that each emotion exhibits a range of styles and intensities, reflecting the natural variability in emotional expression. By analyzing the number of modeled instances, we find that the spherical vectors representing each emotion tend to cluster around specific styles that are frequently associated with that emotion. This suggests that certain emotion styles are more commonly expressed in real-world speech. For example, in the case of surprise, a higher pitch is more frequently observed than a lower pitch. However, for styles with fewer data instances, the distribution showed less distinct patterns, likely due to the limited availability of data. In summary, the results support the assumption that speech emotion analysis should account for diverse styles based on primary emotional states.

\begin{table*}[!ht]
    \centering 
        \caption{Comparison with the results of control methods for non-parallel emotion transfer on the seen dataset. \\ The nMOS, sMOS, and eMOS scores are presented with 95\% confidence intervals.}
    \label{Table4}\vspace{-0.2cm}
        \resizebox{1.00\textwidth}{!}{
    \begin{tabular}{l|ccc|ccc|ccc|ccc}
        \toprule
        \multirow{2}{*}{\textbf{Method}} & \multicolumn{3}{c|}{\textbf{Subjective Evaluation}} & \multicolumn{8}{c}{\textbf{Objective Evaluation}} \\ 
        \cmidrule{2-13} 
         & \textbf{nMOS} ($\uparrow$) & \textbf{sMOS} ($\uparrow$) & \textbf{eMOS} ($\uparrow$) & \textbf{WER\textsubscript{Whis}} ($\downarrow$) & \textbf{WER\textsubscript{w2v}} ($\downarrow$) & \textbf{WER\textsubscript{AVG}} ($\downarrow$) & \textbf{SECS\textsubscript{R}} ($\uparrow$) & \textbf{SECS\textsubscript{W}} ($\uparrow$) & \textbf{SECS\textsubscript{AVG}} ($\uparrow$) & \textbf{ECA} ($\uparrow$) & \textbf{SVAS} ($\uparrow$) & \textbf{EECS} ($\uparrow$) \\ 
        \midrule
            GT & 4.06$\pm$0.05 & 4.15$\pm$0.05 & 4.11$\pm$0.05 & 11.42 & 14.99 & 13.21 & 0.7358 & 0.9264 & 0.8311 & 95.53 & - & 0.9487 \\ 
            BigVGAN \cite{lee2023bigvgan} & 3.95$\pm$0.05 & 4.01$\pm$0.06 & 3.98$\pm$0.06 & 11.36 & 15.15 & 13.26 & 0.7271 & 0.9231 & 0.8251 & 94.25 & 0.9815 & 0.9389 \\ 
        \midrule
            MatchaTTS w/ Scaling Factor \cite{li2022cross} & 3.81$\pm$0.06 & 3.95$\pm$0.06 & 3.82$\pm$0.07 & 18.36 & 23.57 & 20.97 & 0.7212 & 0.9012 & 0.8112 & 40.63 & 0.8191 & 0.4757 \\
            MatchaTTS w/ Relative Attribute \cite{zhou2022emotion} & 3.86$\pm$0.06 & 3.94$\pm$0.06 & 3.84$\pm$0.07 & 15.53 & 19.87 & 17.70 & \textbf{0.7335} & \textbf{0.9066} & \textbf{0.8201} & 85.31 & 0.8672 & 0.8756 \\
        \midrule
            EmoSphere ++ (Proposed) & \textbf{3.92$\pm$0.06} & \textbf{3.97$\pm$0.06} & \textbf{3.86$\pm$0.06} & \textbf{15.52} & \textbf{18.85} & \textbf{17.19} & 0.7314 & 0.9047 & 0.8181 & \textbf{93.53} & \textbf{0.8717} & \textbf{0.9270}\\ 
        \bottomrule
    \end{tabular}
      }\vspace{0cm}
\end{table*}

\begin{figure*}[!t] 
    \centering
\includegraphics[width=1\linewidth]{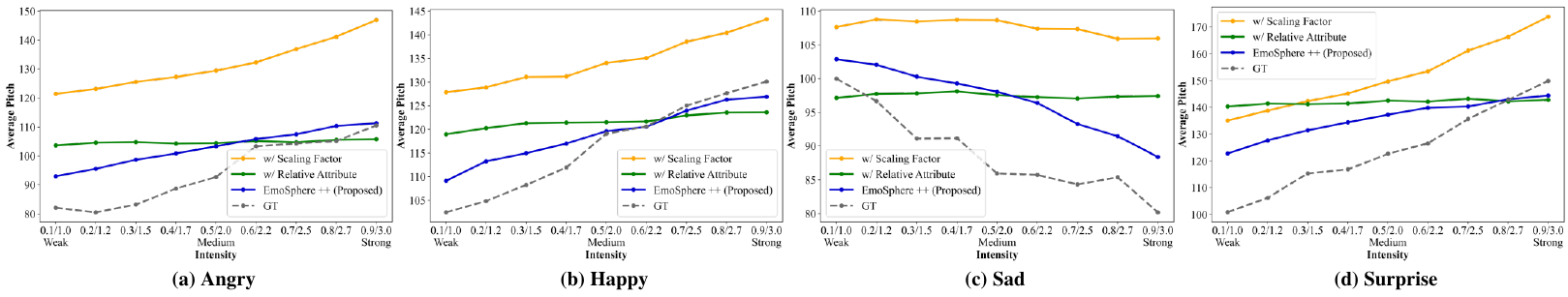}\vspace{-0.2cm}
\caption{Pitch tendency track according to intensity for different emotions. Pitch values were calculated by averaging the synthesized speech for each intensity across all test sentences.
Since the intensity of ground truth (GT) speech cannot be adjusted, the GT line represents the pitch tendency based on the emotion-adaptive spherical vector intensity labels across all test sentences, serving as a reference guideline.
}
\label{Intensity_control}\vspace{0cm}
\end{figure*}

\begin{figure}[!t] 
    \centering 
    \includegraphics[width=0.65\linewidth]{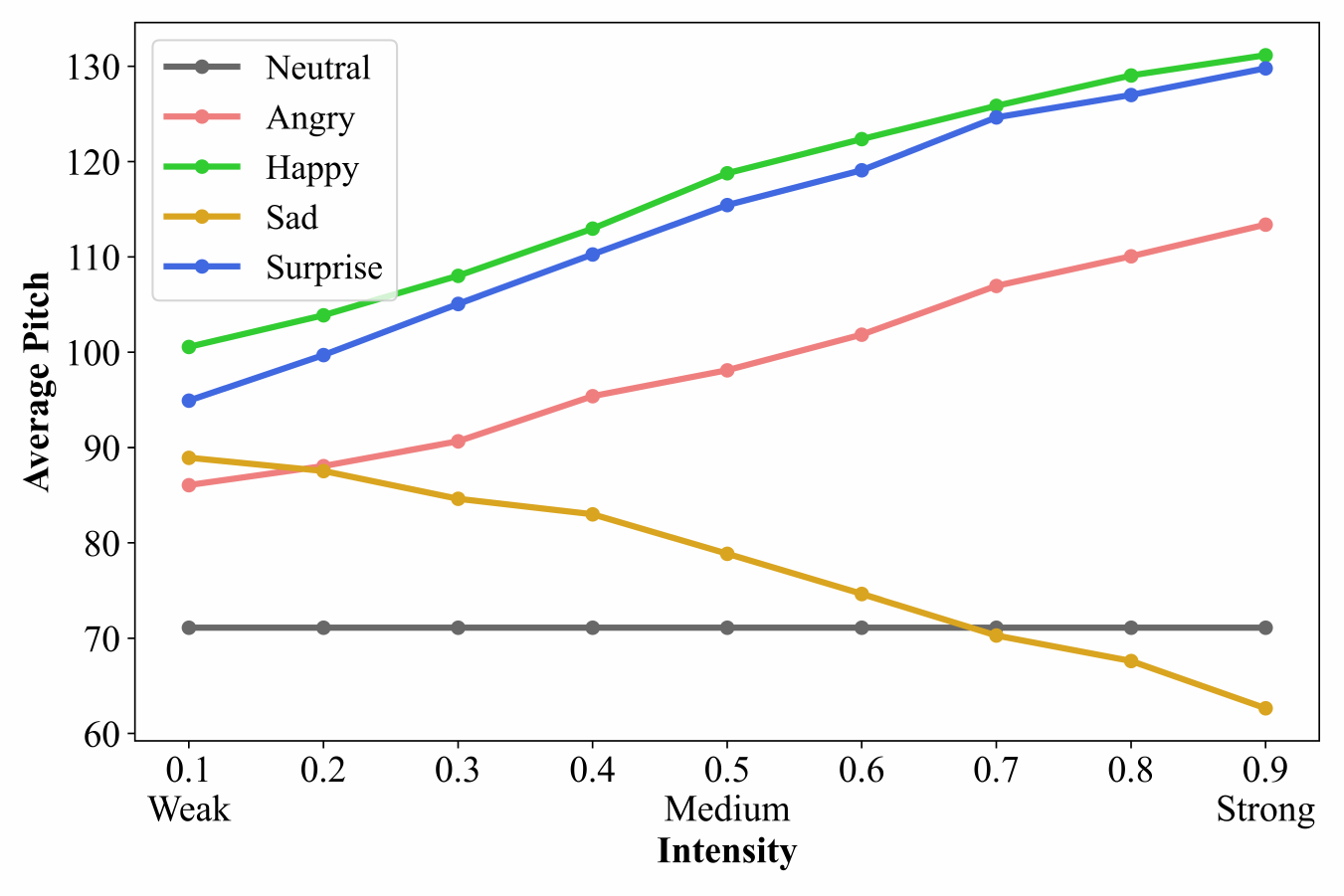}\vspace{-0.2cm}
    \caption{Pitch tendency track according to intensity for an unseen speaker.}
\label{Unseen_intensity_control}\vspace{0cm}
\end{figure}

\subsubsection{Prosodic Variation with Intensity Based on the Valence-Arousal-Dominance (VAD) Axis}
In this section, we describe the analysis of how prosodic variation varies with emotion intensity based on the VAD axis. We represent the region divided into thirds by intensity with Q1, Q2, and Q3, using 0.33 and 0.66 as thresholds. In psychology \cite{russell1980circumplex, russell1977evidence}, valence represents the positivity or negativity of an emotion, arousal indicates the intensity of the emotion provoked by a stimulus, and dominance denotes the level of control exerted by the stimulus. Building on this, we show how prosodic variations reflect diverse emotions through emotion style and intensity along the VAD axis.

Existing studies \cite{inoue2024hierarchical} on emotion control have demonstrated the effectiveness of analyzing prosodic features such as pitch, energy, and duration. Expanding on this approach, we conducted a prosodic analysis and found that: 1) positive valence is associated with higher prosodic feature value as the emotional intensity increases. 2) positive arousal leads to increased patterns of prosodic feature value change. 3) positive dominance results in a narrower prosodic variation range. To further analyze and clearly illustrate this prosodic variation, we introduced three key elements in our analysis. First, to compare differences based on valence, we calculated the average prosodic feature values \textit{AVG.} for each emotion style. Second, to observe increases and decreases related to arousal, we highlighted the highest and lowest prosodic feature values within each intensity level using green and red colors, respectively. Lastly, to examine the variation ranges influenced by dominance, we introduced $Qc$, which represents the absolute difference between the highest and lowest prosodic feature values. These measures provide a more comprehensive understanding of how prosodic features vary with emotional style and intensity.

To validate these findings, we examined specific cases where emotion styles differ along a single VAD dimension while remaining constant in the others. First, when comparing styles ``III'' and ``IV'', where only valence differs, we observe that positive valence is associated with higher average prosodic feature values \textit{AVG.}. Specifically, across all emotions, converting to positive valence results in an average increase of 13.45 in pitch, 1.4 in energy, and 0.25 in duration. Similarly, when comparing styles ``I'' and ``V'', where only dominance differs, we find that $Qc$ is smaller in the positive dominance condition. Specifically, converting to positive dominance results in an average decrease of 6.78 in pitch, 0.35 in energy, and 0.02 in duration across all emotions. Lastly, the positive arousal style of ``I'' exhibits increasing prosodic patterns, whereas the negative arousal style of ``IV'' shows decreasing patterns. These results show that prosodic variations align with the VAD characterization in most cases, supporting the effectiveness of EASV in modeling emotion style and intensity.

\subsection{Model Performance}
We conducted experiments including seen and unseen speaker scenarios to evaluate EmoSphere++ and baseline models for non-parallel style transfer. We split our experiments into two categories: 1) seen non-parallel style transfer and 2) unseen non-parallel style transfer.

\begin{figure}[!t] 
    \centering 
    \includegraphics[width=1.0\linewidth]{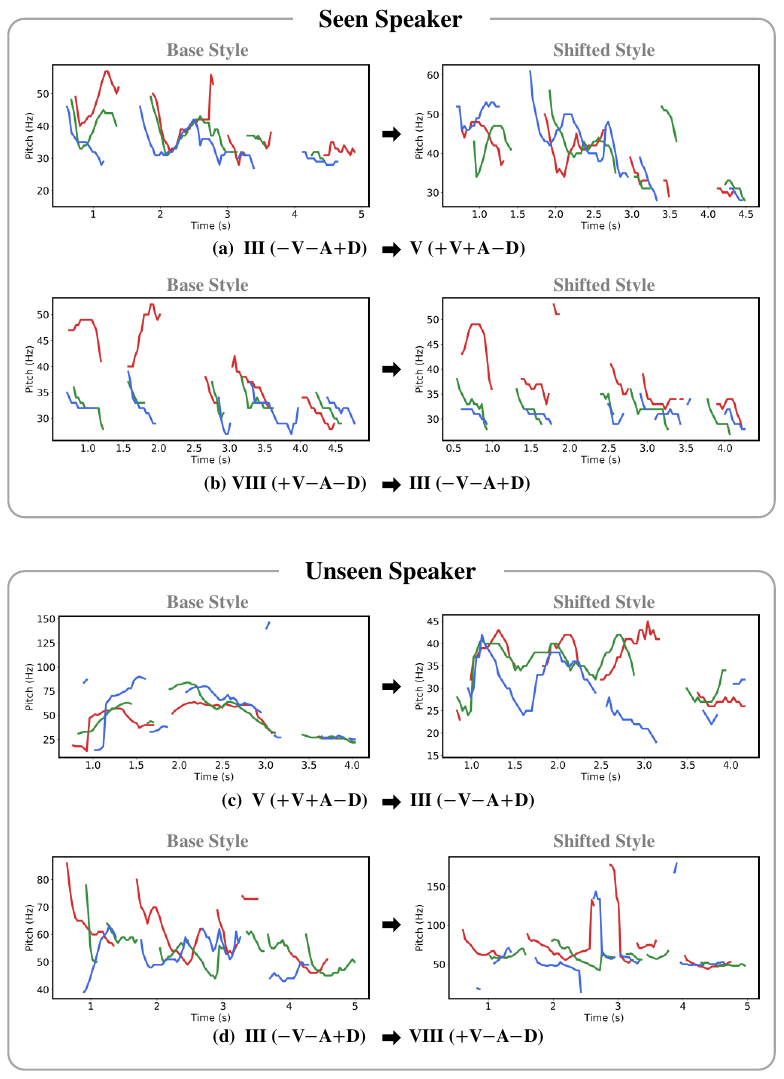}\vspace{-0.2cm}
    \caption{Pitch tracks of a sample demonstrating the effects of emotional style shift in sad emotion, where $A$, $V$, and $D$ represent arousal, valence, and dominance, respectively. The line color represents emotional intensity: red = 0.1, green = 0.5, and blue = 0.9.}
\label{style_shift_sample}\vspace{-0.0cm}
\end{figure}

\subsubsection{Seen Non-Parallel Style Transfer}
We first demonstrate the robustness of our proposed model in seen non-parallel style transfer, where a TTS system synthesizes a seen speaker. We select another fixed pair of reference signals from the test set of the same emotion and speaker. As shown in Table \ref{Table2}, EmoSphere++ outperforms the previous methods in terms of both subjective and objective evaluations. In terms of naturalness and linguistic consistency, EmoSphere++ achieves the highest nMOS with a score of 3.92 and performed strongly in WER with a score of 15.52 compared to those of the baseline models. Regarding emotion and speaker style similarity, EmoSphere++ scores the highest overall eMOS of 3.86 and sMOS of 3.97. The objective results of the speaker metric SECS and emotion metrics ECA, SVAS, and EECS show that EmoSphere++ outperforms state-of-the-art models in transferring custom speech styles. The proposed method demonstrates superior performance in style transfer and quality compared to previous approaches.

\subsubsection{Unseen Non-Parallel Style Transfer}
Subsequently, we explored the robustness of the proposed model in the case of unseen non-parallel style transfer. We set up zero-shot scenarios with unseen speakers for the ESD dataset and tested how the TTS model reproduces each speaker style when synthesizing different emotional phrases. Mellotron \cite{valle2020mellotron} and Mixedemotion \cite{zhou2023speech} are not suitable for zero-shot scenarios because they rely on speaker lookup tables and are fine-tuned to a single speaker. As shown in Table \ref{Table3}, the result indicates the drop in overall metrics, indicating that adapting to unseen speakers is more complex than adapting to unseen emotions. However, the most consistent performance in the various emotion zero-shot scenarios suggests that EmoSphere++ can adjust to a wide range of unseen emotions.

\begin{figure}[!t] 
    \centering
\includegraphics[width=0.9\linewidth]{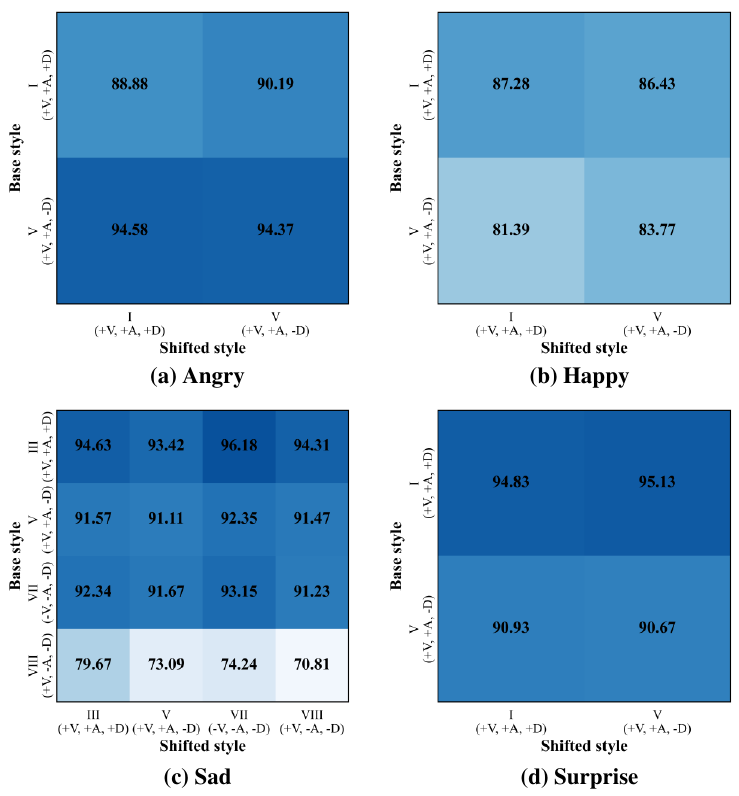}\vspace{-0.2cm}
\caption{Comparison of the emotion classification accuracy scores of shifting emotion style.}
\label{Style_control_ECA}\vspace{-0.0cm}
\end{figure}

\subsection{Emotion Intensity Control}
\subsubsection{Comparison With Control Methods}
As a comparative study, we implemented two speaker and emotion attribute control methods (i.e., Matcha-TTS w/ Scaling Factor, Matcha-TTS w/ Relative Attributes), as described in Section \ref{Comparison Models}. We evaluated the performance of these two methods in terms of performance and control. To demonstrate the ability of our model to control intensity, we synthesized speech with three different levels of emotion intensity (weak, medium, and strong). In the relative attribute model and EmoSphere++, we uniformly refer to values of 0.1 as weak, 0.5 as medium, and 0.9 as strong. The scaling factor cannot assign intensity values; therefore, we set the scalar factor at 1, 2, and 3 to represent weak, medium, and strong emotion intensities, as in the original setting.
The GT line reflects the pitch tendency of the original speech, based on inherent EASV intensity labels across all test sentences, serving as a reference guideline.

\textbf{Performance}. We explored the expressiveness of our proposed model in non-parallel style transfer, where a TTS system synthesizes both the prosodic style of a reference signal and modeling emotion attribute. The results are compiled and presented in Table \ref{Table4} for easy comparison. Compared to transferring emotion from label-based emotion with relative attributes and reference-based emotion with a scaling factor, improved speech quality and expressiveness are achieved using EASV.

\textbf{Control}.  We calculated the average pitch of the synthesized speech across all test sentences based on the intensity of each emotion. Fig. \ref{Intensity_control} shows the following: 1) The Matcha-TTS w/ Relative Attributes reflects the adjustment of emotional properties, but also often shows a limited adjustable range and tends to be reduced to a more uniform style. This outcome suggests subtle emotional nuances cannot be easily captured using only emotion labels. 2) The Matcha-TTS w/ Scaling Factor exhibits the most variation, but often becomes unstable when adjusted on labels such as sad. Therefore, determining an appropriate scaling factor is difficult, and adjustments may lead to instability in audio quality. Conversely, the pitch tendency plot of EmoSphere++ closely follows the GT line, reflecting intensity variations based on emotion. This result indicates that the proposed model synthesizes speech according to the given intensity scale while effectively capturing variations that align more closely with natural emotional speech patterns.

\begin{table*}[t]
    \centering
    \caption{
    Ablation study on disentangling method and joint attribute style encoder for non-parallel emotion transfer on the seen and unseen datasets. 
    The nMOS, sMOS, and eMOS scores are presented with 95\% confidence intervals.
    }\vspace{-0.2cm}
    \resizebox{1.00\textwidth}{!}{
        \begin{tabular}{l|ccc|c|cc|ccc|c|cc}
        \toprule
            \multirow{3}{*}{\textbf{Method}} & \multicolumn{6}{c|}{\textbf{Seen Speaker}} & \multicolumn{6}{c}{\textbf{Unseen Speaker}} \\ 
        \cmidrule{2-13} 
             & \multicolumn{3}{c|}{\textbf{Subjective Evaluation}} & \multicolumn{3}{c|}{\textbf{Objective Evaluation}} & \multicolumn{3}{c|}{\textbf{Subjective Evaluation}} & \multicolumn{3}{c}{\textbf{Objective Evaluation}} \\ 
        \cmidrule{2-13} 
            & \textbf{nMOS} ($\uparrow$) & \textbf{sMOS} ($\uparrow$) & \textbf{eMOS} ($\uparrow$) & \textbf{SECS\textsubscript{AVG}} ($\uparrow$) & \textbf{ECA} ($\uparrow$) & \textbf{EECS} ($\uparrow$) & \textbf{nMOS} ($\uparrow$) & \textbf{sMOS} ($\uparrow$) & \textbf{eMOS} ($\uparrow$) & \textbf{SECS\textsubscript{AVG}} ($\uparrow$) & \textbf{ECA} ($\uparrow$) & \textbf{EECS} ($\uparrow$) \\ 
        \midrule
            GT & 4.23$\pm$0.04 & 4.19$\pm$0.03 & 4.08$\pm$0.03 & 0.8311 & 95.53 & 0.9487 & 4.16$\pm$0.05 & 4.11$\pm$0.05 & 4.07$\pm$0.03 & 0.8504 & 100.00 & 0.9944 \\
            BigVGAN & 4.22$\pm$0.04 & 4.13$\pm$0.04 & 4.05$\pm$0.04 & 0.8251 & 94.25 & 0.9389 & 4.12$\pm$0.06 & 4.05$\pm$0.04 & 4.02$\pm$0.04 & 0.8439 & 99.32 & 0.9903 \\
        \midrule
            w/o Disentangling Method & 3.83$\pm$0.05 & \textbf{3.78$\pm$0.05} & 3.75$\pm$0.04 & 0.8171 & 92.86 & 0.9218 & 3.70$\pm$0.08 & \textbf{3.83$\pm$0.05} & 3.70$\pm$0.04 & 0.7573 & 88.56 & 0.8802 \\
             w/o Dimensional Emotion Encoder  & 3.79$\pm$0.05 & 3.75$\pm$0.04 & 3.74$\pm$0.05 & 0.8165 & 92.18 & 0.9272 & 3.73$\pm$0.08 & 3.81$\pm$0.04 & 3.72$\pm$0.05 & 0.7574 & 88.27 & 0.8900 \\
             w/o Global Emotion Encoder  & 3.76$\pm$0.05 & 3.77$\pm$0.04 & 3.51$\pm$0.05 & 0.8178 & 77.68 & 0.8171 & 3.64$\pm$0.08 & 3.79$\pm$0.05 & 3.60$\pm$0.04 & 0.7568 & 59.66 & 0.6712  \\
        \midrule
             w/ Gradient Reversal Layer \cite{oh2024durflex}  & 3.82$\pm$0.05 & 3.76$\pm$0.04 & 3.77$\pm$0.04 & 0.8169 & 92.76 & \textbf{0.9283} & 3.75$\pm$0.07 & 3.81$\pm$0.05 & 3.71$\pm$0.04 & 0.7566 & 89.58 & 0.8947 \\
            w/ Vector Quantization \cite{zhang2023iemotts} & 3.78$\pm$0.05 & 3.74$\pm$0.03 & 3.76$\pm$0.05 & 0.8143 & 92.34 & 0.9235 & \textbf{3.76$\pm$0.08} & 3.67$\pm$0.05 & 3.74$\pm$0.04 & 0.7262 & 92.60 & 0.9320 \\
            w/ Orthogonality Loss \cite{li2022cross} & 3.79$\pm$0.06 & 3.76$\pm$0.04 & 3.76$\pm$0.04 & 0.8160 & 92.26 & 0.9191 & 3.74$\pm$0.08 & 3.74$\pm$0.05 & 3.73$\pm$0.05 & 0.7549 & 91.23 & 0.9101 \\
        \midrule
            EmoSphere ++ (Proposed) & \textbf{3.87$\pm$0.05} & 3.77$\pm$0.04 & \textbf{3.78$\pm$0.05} & \textbf{0.8181} & \textbf{93.53} & 0.9270 & \textbf{3.76$\pm$0.07} & \textbf{3.83$\pm$0.05} & \textbf{3.75$\pm$0.04} & \textbf{0.7592} & \textbf{94.61} & \textbf{0.9385} \\
        \bottomrule
        \end{tabular}
    }
    \label{Table5} \vspace{0cm}
\end{table*}

\subsubsection{Intensity Control in the Zero-Shot Scenario}
To demonstrate the ability to control emotional expression in the zero-shot scenario, we visualized the tendency of the pitch as shown in Fig. \ref{Unseen_intensity_control}. We calculated the average pitch of the synthesized speech for the unseen speaker across all test sentences based on the intensity of each emotion. The pitch trend graph changes with intensity, reflecting the nature of the emotion. We observe a decrease in pitch for the sad emotion, whereas the pitch tends to increase as intensity rises for other emotions. This pattern indicates that the synthesized speech in EmoSphere++ can control the intensity of each emotion, even for the zero-shot scenario.

\subsection{Emotion Style Shift}
\subsubsection{Visual Comparisons}
We visualized the prosodic attributes of pitch related to the emotion intensity to gain an intuitive understanding of emotion style. To illustrate the variation patterns in emotion intensity with shifted emotion style, we visualized the pitch track changes for a sample, both from seen and unseen speakers. As analyzed in Section \ref{Analysis of Emotional Style and Intensity Modeling}, the prosodic pattern based on emotional intensity reflects the characteristics of the VAD axis. As shown in Fig. \ref{style_shift_sample}, we visualized the pitch contour of sad utterances with the most varied styles. For example, style vectors with positive V axes have higher average pitch values and reduced duration; positive A axes have a pitch that tends to increase the changing patterns. By contrast, duration decreases, and positive D axes have a narrow range in changing patterns with a broader duration. These results indicate that the proposed EASV is meaningfully characterized and enables emotion style and intensity controllable speech synthesis in both seen and unseen.

\subsubsection{Comparison of Emotional Consistency}
Fig. \ref{Style_control_ECA} shows the resulting ECA for representative combinations of style shift. Across all emotions, we observe a similar emotional consistency when shifting to the representative style combinations as when maintaining the original style. These results indicate that the proposed model effectively adjusts emotion styles while maintaining emotional consistency across all style transformations. Therefore, as hypothesized in various psychological theories \cite{plutchik2013theories, reisenzein1994pleasure}, spherical vectors of emotion style can be characterized as derivatives of basic emotions.

\subsection{Ablation Study}
\subsubsection{Impact of Disentangling Method and Joint Attribute Style Encoder}
As shown in Table \ref{Table5}, we conducted ablation studies to evaluate the impact of the disentangling method and joint attribute style encoder. To ensure a fair comparison of the proposed joint attribute style encoder modules and the existing disentangling approaches, we trained the models using the same dataset and configurations as EmoSphere++.  We replaced the existing emotion disentangling method with three competing disentangling approaches: 1) \textbf{w/ Gradient Reversal Layer} \cite{oh2024durflex}, which employs adversarial speaker training using the GRL \cite{ganin2016domain} applied after the fully connected layer that follows the global encoder; 2) \textbf{w/ Vector Quantization} \cite{zhang2023iemotts}, which implements a bottleneck layer via a modified VQ layer \cite{van2017neural} applied after the fully connected layer that follows the global encoder; and 3) \textbf{w/ Orthogonality Loss} \cite{li2022cross}, which constrains the emotion embedding and speaker embedding using an orthogonal loss \cite{ranasinghe2021orthogonal} without normalizing all sample pairs. To further evaluate the effectiveness of the joint attribute style encoder, we conducted additional ablation studies by removing individual encoder components: \textbf{w/o Global Emotion Encoder}, where the global emotion embedding is excluded and \textbf{w/o Dimensional Emotion Encoder}, where the dimensional-driven emotion embedding is removed. Additionally, to analyze the role of the disentangling method, we trained a variant of the proposed model without the disentangling method, referred to as \textbf{w/o Disentangling Method}. In the \textbf{w/o Global Emotion Encoder} setting, the disentangling method was modified to use the dimensional-driven emotion embedding in place of the global emotion embedding.

\begin{table*}[!ht]
    \centering 
        \caption{Ablation study on VAD extractor for parallel emotion transfer on IEMOCAP datasets.}
    \label{Table6}\vspace{-0.2cm}
        \resizebox{1.00\textwidth}{!}{
    \begin{tabular}{l|ccc|ccc|ccc|ccc}
        \toprule
        \textbf{Method} & \textbf{WER\textsubscript{Whis}} ($\downarrow$) & \textbf{WER\textsubscript{w2v}} ($\downarrow$) & \textbf{WER\textsubscript{AVG}} ($\downarrow$) & \textbf{SECS\textsubscript{R}} ($\uparrow$) & \textbf{SECS\textsubscript{W}} ($\uparrow$) & \textbf{SECS\textsubscript{AVG}} ($\uparrow$) & $\textbf{RMSE}_{f0}$ ($\downarrow$) & $\textbf{RMSE}_{period}$ ($\downarrow$) & \textbf{F1 {V/UV}} ($\uparrow$) & \textbf{ECA} ($\uparrow$) & \textbf{SVAS} ($\uparrow$) & \textbf{EECS} ($\uparrow$) \\ 
        \midrule
            GT & 13.70 & 28.24 & 20.97 & 0.7383 & 0.8506 & 0.7945 & - & - & - & 54.50 & - & 0.6444 \\ 
            BigVGAN & 14.93 & 13.95 & 14.44 & 0.6449 & 0.8496 & 0.7473 & 3.02 & 0.3262 & 0.6070 & 53.51 & 0.9685 & 0.6338 \\ 
        \midrule
            w/ Real VAD Value & 28.51 & \textbf{46.25} & 37.38 & \textbf{0.6153} & 0.8283 & 0.7218 & 20.17 & \textbf{0.4580} & 0.4971 & \textbf{40.34} & 0.8407 & 0.5213 \\ 
            w/ Predict VAD Value (Proposed) & \textbf{27.07} & 46.53 & \textbf{36.80} & 0.6149 & \textbf{0.8336} & \textbf{0.7243} & \textbf{19.95} & 0.4587 & \textbf{0.4997} & 39.50 & \textbf{0.8505} & \textbf{0.5450} \\
        \bottomrule
    \end{tabular}
      }\vspace{0cm}
\end{table*}

\begin{figure*}[!t] 
    \centering
    \includegraphics[width=1.0\linewidth]{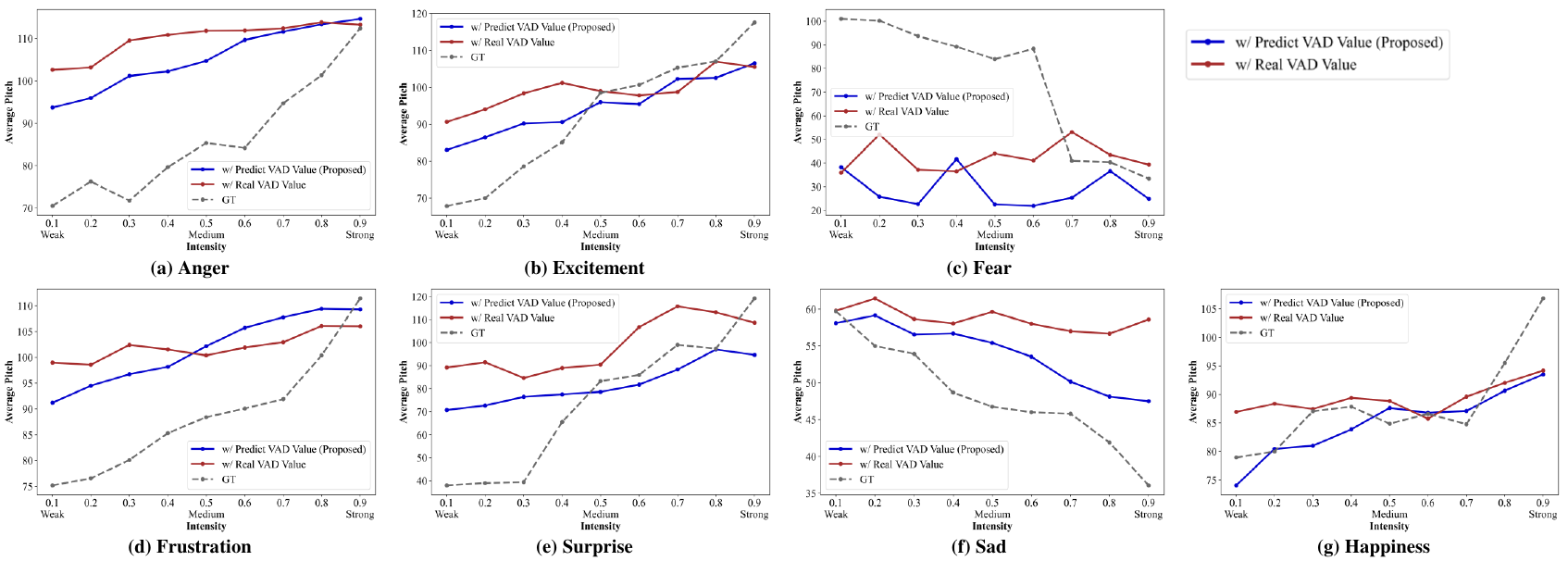}\vspace{-0.2cm}
    \caption{
    Pitch tendency track according to intensity for each emotion. Pitch values were calculated by averaging the synthesized speech for each intensity across all test sentences.
    Since the intensity of ground truth (GT) speech cannot be adjusted, the GT line represents the pitch tendency based on the emotion-adaptive spherical vector intensity labels across all test sentences, serving as a reference guideline.
}\vspace{0cm}
    \label{iemocap_intensity_control}
\end{figure*}

We used comparative subjective metrics (nMOS, sMOS, and sMOS) and objective metrics (SECS, ECA, and EECS) to assess the expressiveness and quality of the generated speech. 
The experimental results show that improvements in both quality and expressiveness highlight the effectiveness of disentangling emotion and speaker embeddings, ensuring clearer emotional transfer without compromising speaker identity. Furthermore, our findings demonstrate that integrating global and dimensional-driven features in the joint attribute style encoder enables the model to capture both broad and fine-grained characteristics, further enhancing expressive. Specifically, the proposed normalized orthogonality loss is crucial for preserving emotional expressiveness in unseen cases, reinforcing its importance for achieving strong generalization performance in zero-shot scenarios.

\subsubsection{Effectiveness of Predicted versus Real VAD Values}
To compare the impact of emotional attribute prediction models \cite{wagner2023dawn}, we conducted comparative experiments using the IEMOCAP dataset \cite{busso2008iemocap}, which includes emotion dimension annotations. As a training input for the emotion-adaptive coordinate transformation, we compared two features: real VAD values from human labeled annotations and predicted VAD values from the emotional attribute prediction \cite{wagner2023dawn}. 
Table \ref{Table6} shows that the objective metrics of prosodic expressiveness for both real and predicted VAD values are similar, suggesting that the predicted values perform comparably to manual annotations. 
Additionally, as shown in Fig. \ref{iemocap_intensity_control}, both the real and predicted VAD values exhibit a consistent and accurate pattern across all emotions, aligning well with the ground truth in intensity control. 
The results indicate that the VAD values predicted from the emotional attribute prediction \cite{wagner2023dawn} provide reliable emotion style and intensity modeling comparable to the real VAD values.

\subsubsection{Comparison of Coordinate Transformation}
\begin{figure}[!t] 
    \centering 
    \includegraphics[width=1\linewidth]{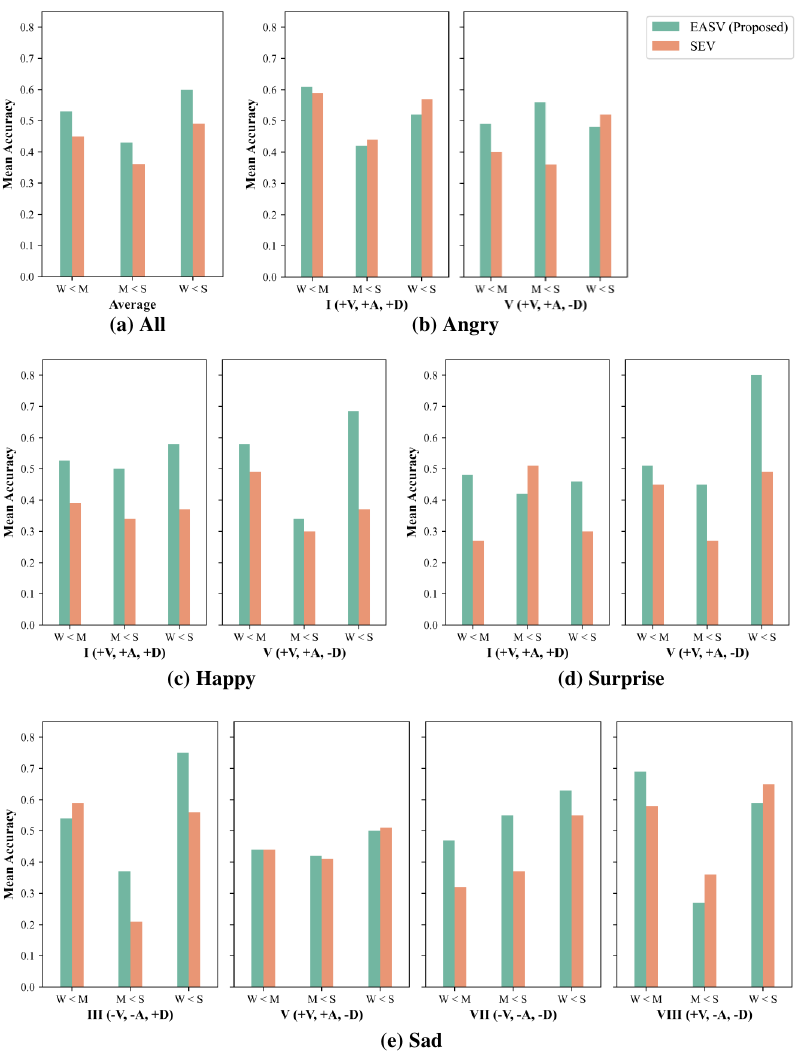}\vspace{-0.2cm}
    \caption{Evaluation process rates the discriminability of synthesized speech samples across different intensity levels, with $W$, $M$, and $S$ representing weak, medium, and strong intensities, respectively.}
\label{intensity_preference} \vspace{0cm}
\end{figure}

For a fair comparison, we compared intensity accuracy using speech pairs generated from the same model, with identical emotion and style, differing only in their coordinate transformation approach. As shown in Fig. \ref{intensity_preference}, all types of emotion intensity pairs (W$<$M, M$<$S, W$<$S) demonstrate accuracy across individual emotions, styles, and overall. We referred to values of W as weak, M as medium, and S as strong. This experiment was conducted to evaluate whether more precise modeling of emotion style and intensity leads to clearly distinguishable synthesized speech when given pseudo-labels as input. By assessing the accuracy of intensity differentiation, we aim to validate the effectiveness of EASV in generating perceptually distinct emotional variations. The EASV method demonstrates consistently high average accuracy for individual emotions and overall. These results suggest that, while the mean-based approach of SEV captures the emotion to a certain extent, it remains unstable at specific intensity levels and styles. This indicates that considering the distribution of other emotional categories in EASV further improves the modeling of emotion style and intensity.

\section{Discussion}
This study represents an initial attempt at modeling and synthesizing emotion styles and intensities for controllable emotional speech synthesis. Although we have demonstrated the effectiveness of our method, some related issues remain unresolved. We discuss these issues and aim to inspire future studies.

\subsection{Limitations of Data Imbalance}
As mentioned in Section \ref{Analysis of Emotional Style and Intensity Modeling}, the emotional styles modeled in each spherical vector exhibit unique characteristics that are confined to specific emotions. These results indicate that people tend to express particular emotional styles more frequently in reality; therefore, we focused on using only representative styles. This simplification may pose a limitation in capturing the full range of emotional nuances. However, researchers can address this issue by expanding the model to include more diverse datasets. Moreover, this serves as the initial attempt to model and synthesize emotion styles and intensities, demonstrating the potential of this approach.

\subsection{Remaining Challenges of Emotional Attribute Prediction Model}
As we summarized the emotional attribute prediction models in Section \ref{Emotion-Adaptive Coordinate Transformation}, we utilized the fine-tuned wav2vec 2.0 for an emotional attribute prediction \cite{wagner2023dawn} task. In the study, VAD predicted by the wav2vec 2.0-based emotional attribute prediction model \cite{wagner2023dawn} was significant. Hence, we utilized VAD pseudo-labels to avoid the inherent subjectivity and high costs associated with data collection. However, the approach relies on the performance of the trained emotional attribute prediction model and contains similar biases and challenges to those encountered in emotional attribute prediction \cite{wagner2023dawn}. We expect the extended VAD predictor \cite{zhou2024emotional} to mitigate the biases inherent in emotional attribute prediction while providing more accurate VAD estimations, ultimately addressing this issue.

\section{Conclusion}
In this paper, we presented EmoSphere++, an emotion-controllable zero-shot TTS model that can control emotional style and intensity to resemble natural human speech. To achieve this, we proposed the novel emotion-adaptive spherical vector (EASV), which models emotional style and intensity as derivatives of primary emotions. Building on this, our comprehensive analysis validates that a dimensional model can characterize emotional states in relation to primary emotions. Additionally, a zero-shot speech synthesis framework with rich expressiveness and controllability was developed, utilizing a joint attribute style encoder with additional loss functions, without being restricted by predefined speaker and emotion labels. The experimental results thoroughly analyze the components of our model and demonstrate its ability to effectively synthesize and control speech performance, even in zero-shot scenarios. We demonstrated that by controlling the spherical vector along the VAD axis, explicit adjustments to emotional style and intensity enable fine-grained emotional expression. Ablation studies further confirm that the proposed EASV effectively synthesizes the complex nature of emotion. While this article only focused on studying emotion-controllable TTS for a limited set of emotions, our proposed spherical vector can enable complete emotion control in most existing emotional speech synthesis frameworks. Future work will expand these experiments to include emotional voice conversion.
\bibliographystyle{IEEEtran}
\bibliography{refs}
\vskip -2\baselineskip plus -1fil
\begin{IEEEbiography}[{\includegraphics[width=1in,height=1.25in,clip,keepaspectratio]{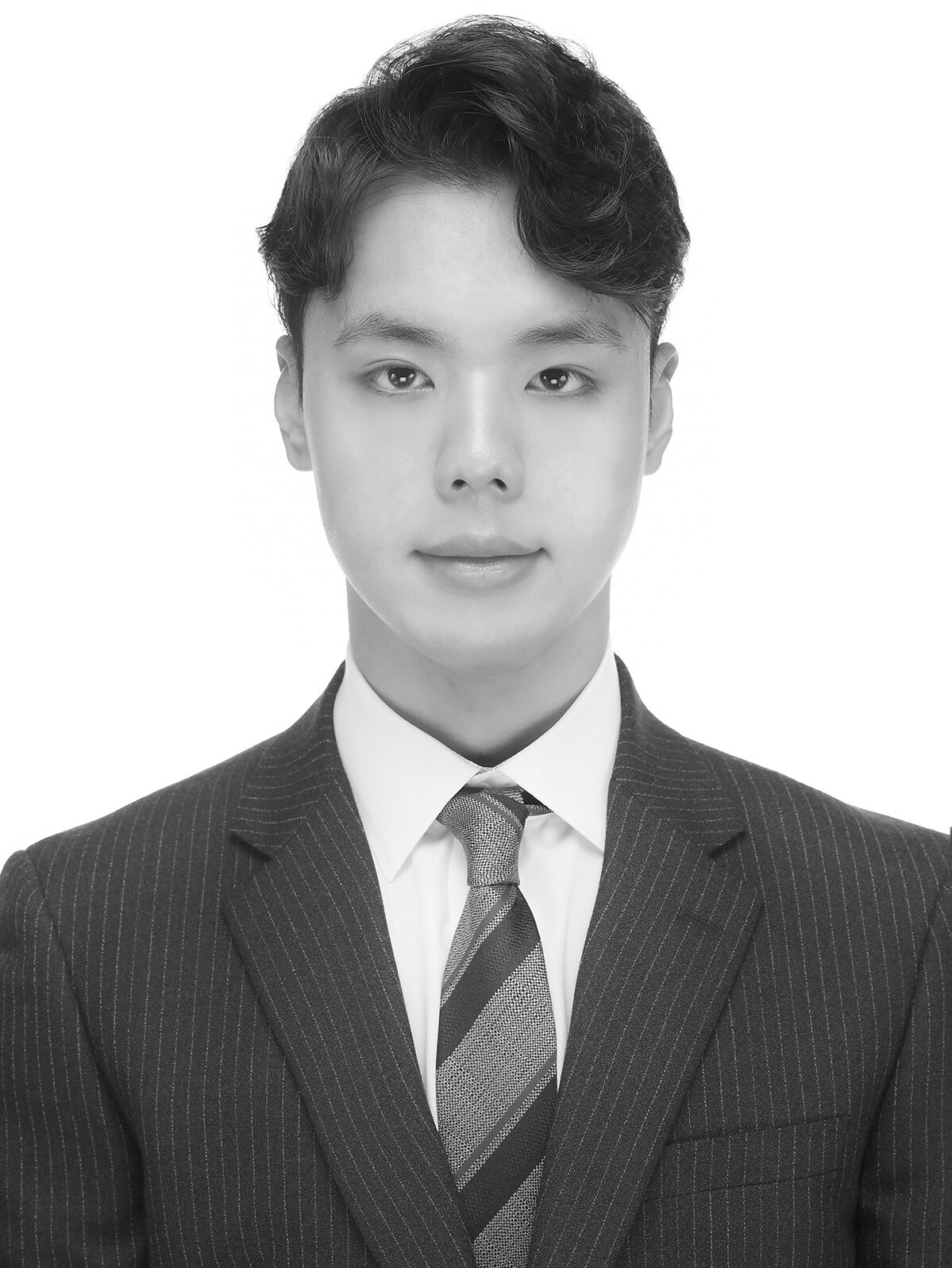}}]{Deok-Hyeon~Cho}
received the B.S. degree in Applied Mathematics from Hanyang University ERICA Campus, Ansan, South Korea, in 2022. He is currently working toward an integrated master's and Ph.D. degree with the Department of Artificial Intelligence, Korea University, Seoul, South Korea. His research interests include artificial intelligence and audio signal processing.
\end{IEEEbiography}

\vskip -2\baselineskip plus -1fil
\begin{IEEEbiography}[
{\includegraphics[width=1in,height=1.25in,clip,keepaspectratio]{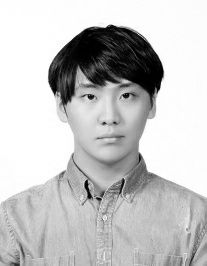}}]{Hyung-Seok~Oh}
received the B.S. degree in Computer Science and Engineering from Konkuk University, Seoul, South Korea, in 2021. 
He is currently working toward an integrated master's and Ph.D. degree with the Department of Artificial Intelligence, Korea University, Seoul, South Korea. 
His research interests include artificial intelligence and audio signal processing.
\end{IEEEbiography}

\vskip -2\baselineskip plus -1fil
\begin{IEEEbiography}[
{\includegraphics[width=1in,height=1.25in,clip,keepaspectratio]{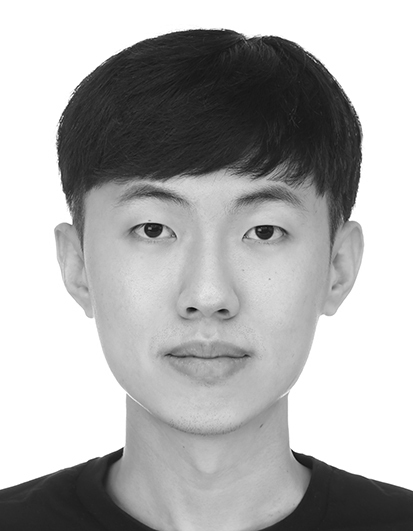}}]{Seung-Bin~Kim}
received the B.S. degree in Physics from University of Seoul, South Korea, in 2021. He is currently pursuing an integrated Master's and Ph.D. degrees with the Department of Artificial Intelligence, Korea University, South Korea. His research interests include artificial intelligence and audio signal processing.
\end{IEEEbiography}

\vskip -2\baselineskip plus -1fil
\begin{IEEEbiography}[{\includegraphics[width=1in,height=1.25in,clip,keepaspectratio]{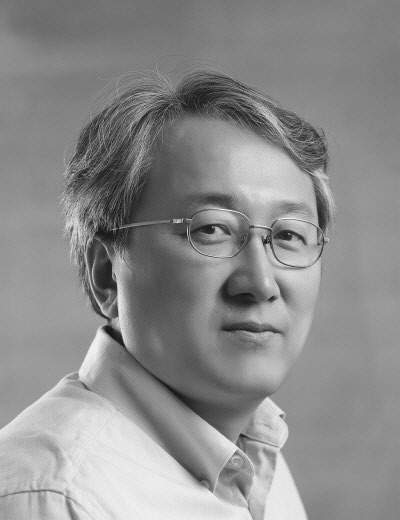}}]{Seong-Whan~Lee}
(Fellow, IEEE) received the B.S. degree in computer science and statistics from Seoul National University, South Korea, in 1984, and the M.S. and Ph.D. degrees in computer science from the Korea Advanced Institute of Science and Technology, South Korea, in 1986 and 1989, respectively. He is currently the Head of the Department of Artificial Intelligence, Korea University, Seoul. His current research interests include artificial intelligence, pattern recognition, and brain engineering. He is a Fellow of the International Association of Pattern Recognition (IAPR), the Korea Academy of Science and Technology, and the National Academy of Engineering of Korea.
\end{IEEEbiography}
\end{document}